\documentclass[12pt,apj,twocolumn]{openjournalb}

\usepackage{amsmath}
\usepackage{lipsum}
\usepackage{multirow}
\usepackage{xcolor}
\usepackage{textgreek}
\usepackage[utf8]{inputenc}
\usepackage[english]{babel}
\usepackage{hyperref}
\hypersetup{
    colorlinks=true,
    linkcolor=blue,
    filecolor=blue,      
    urlcolor=blue,
    citecolor=blue,
}
\usepackage{color,colortbl}
\usepackage{tensind}
\tensordelimiter{?}
\DeclareGraphicsExtensions{.bmp,.png,.jpg,.pdf}
\usepackage{verbatim}
\usepackage[normalem]{ulem}
\usepackage{orcidlink}
\usepackage{soul}

\graphicspath{ {./figs/} }

\def\M1450{M_{\rm 1450}}
\def\CHIMP{~h^{-1}{\rm~Mpc}}

\def\Msun{{\rm M}_\odot}

\def\xHIv{{\left<x_{\rm HI}\right>_{\rm v}}}

\def\ang{\stackrel{\circ}{\rm A}}
\def\gbg{\Gamma_{\rm bkg}}
\def\mfp{\lambda_{\rm  mfp}}
\def\Mag{M_{1450}}
\def\Req{R_{\rm eq}}
\def\B21{\citet{becker2021}}

\begin{document}
\title{Cosmic Reionization On Computers: biases and uncertainties in the measured mean free path at the end stage of reionization}

\author{Huanqing Chen\orcidlink{0000-0002-3211-9642}}
\email{huanqing.chen@ualberta.ca}
\affiliation{Augustana Campus,
University of Alberta,
Camrose, AB T4V2R3, Canada}
\affiliation{Canadian Institute for Theoretical Astrophysics,
University of Toronto,
Toronto, ON M5R 2M8, Canada
}

\author{Jiawen Fan \orcidlink{0009-0002-9477-886X}}
\affiliation{Department of Physics, University of Michigan,  Ann Arbor, MI 48109, USA}

\author{Camille Avestruz\orcidlink{0000-0001-8868-0810}}
\affiliation{Leinweber Center for Theoretical Physics,
}
\affiliation{Department of Physics,
University of Michigan, Ann Arbor, MI 48109, USA
}

\begin{abstract}
Recent observations and analyses of absorption in quasar spectra suggest a rapid drop in the mean free path (MFP) at the late stage of reionization at $z\sim6$.  We use the Cosmic Reionization on Computers simulation to examine potential biases in observed measurements of the MFP at the late stage of reionization, particularly in the presence of a quasar.  
We analyze three snapshots surrounding the `ankle' point of reionization history, when extended neutral patches of the intergalactic medium disappeared in the simulation box. Specifically, these are $z=6.8$ (true MFP $\approx 0.4$~pMpc), in addition to $z=6.1$ (true MFP $\approx 2$~pMpc) and $z=5.4$ (true MFP $\approx 6$~pMpc). 
We compare the inferred MFP $\mfp$ from synthetic spectra fits to the true MFP. We find that the mean Lyman continuum (LyC) profile at $z=6.8$ changes significantly with quasar lifetime $t_Q$. 
We attribute this sensitivity to $t_Q$ to a combination of extended neutral IGM patches and the prevalence of small-scale dense clumps. 
Consequently, the inferred MFP can be biased by a factor of few depending on $t_Q$. 
On the other hand, for the $z=6.1$ and $z=5.4$ snapshots, the mean LyC profile shows minimal sensitivity to variation in $t_Q\gtrsim 1$ Myr. 
The inferred MFP in these two cases is accurate to the $\lesssim 30\%$ level. Our results highlight how modeling systematics can affect the inferred MFP, particularly in the regime of small true MFP ($\lesssim 0.5$ pMpc). We also discuss the potential of this regime to provide a testing ground for constraining quasar lifetimes from LyC profiles.

\end{abstract}


\maketitle

\section{Introduction}
\label{sec:intro}
The epoch of reionization (EoR) marks a crucial transition time of the intergalactic medium (IGM).
During the early stage of the EoR, energetic photons escaped from the first galaxies and ionized the surrounding IGM.
As time elapsed, more galaxies formed and  emitted more energetic photons, enlarging the ionized bubbles in the previously neutral IGM, eventually turning the IGM to a mostly ionized state. The timing and mechanisms of this transition remain subjects of active research.
One key quantity to measure is the mean free path (MFP), which is defined as the average distance ionizing photon can travel before it is absorbed by neutral gas.
Such absorbing neutral gas typically exists in two forms -- extended and attenuated patches, or small and dense clumps known as Lyman limit systems (LLSs) and Damped Lyman alpha systems (DLAs). Therefore, both the merging of ionized bubbles and the rapid photonionization of LLSs/DLAs can lead to a rapid decreasing of the MFP.
One significant progress made in the recent years is the new MFP measurements at $z\approx 6$ \citep{becker2021,zhu2023}. Previous to that, MFP has been measured using the Lyman Continuum (LyC) spectra of quasars but only at $z\lesssim 5$ \citep{prochaska2009,omeara2013,fumagalli2013,worseck2014,lusso2018}. 
The consensus is that the MFP evolves slowly at $2\lesssim z\lesssim 5$ and can be fit by a single power-law across these few gigayears, with a approximated relation MFP$\propto (1+z)^{-5.4}$ \citep{worseck2014}. However, \citet{becker2021,zhu2023} find that between $z=5\sim6$, the relationship deviates significantly from the power-law extrapolation towards a small value of $\mfp \lesssim 1$ pMpc. 
Using a similar formalism but measuring MFP individually, \citet{bosman2021b} obtains a lower-limit of $\mfp \approx 0.3$~pMpc at $z=6$. These are new pieces of evidence supporting the ``late-reionization'' scenario \citep{becker2015,kulkarni2019,keating2020,nasir2020,qin2021,bosman2022}. In particular, \citet{becker2021} finds that when comparing with simulations \citep{keating2020, daloisio2020}, the measured $\mfp$ is even lower than the ``late-reionization'' model where the IGM is still $\sim 20\%$ neutral at $z=6$.


However, measuring the MFP at $z>5$ is much more complicated than that at lower redshifts due to the relative importance of the radiation from the quasar itself. At $z\sim 6$, the the background ionizing radiation is $\gbg \sim 10^{-13}~\rm~ s^{-1}$, as compared to $\gbg \gtrsim 10^{-12} ~\rm~ s^{-1}$ at $z<5$ \citep{gaikwad2023,davies2024}.
Given the much larger luminosity distance at $z\sim 6$, with current facilities, one could only obtain a smaller sample of intrinsically brighter quasars of $\M1450\lesssim -25$. This means that for the sample of quasars used to measure MFP at $z\sim 6$, the radiation from the quasar is larger than the background radiation within the $\sim 10$ pMpc proximity region. This number is an order-of-magnitude larger than the measured $\mfp$. Naturally, this means that to extract the true MFP, one needs to carefully model the radiation from the quasar, otherwise the measurement of MFP could be significantly biased.

To address this issue, \citet{becker2021,zhu2023} adopt a model that accounts for the opacity modification under the quasar radiation. Motivated by analytic models from \citet{miraldaescude2000,furlanetto2005} and radiative transfer simulations in \citet{mcquinn2011}, they formulate the IGM opacity under quasar radiation to be the one without quasar radiation multiplying a factor of $\left(\frac{\Gamma_{\rm qso} + \gbg}{\gbg}\right)^{-\xi}$, {where $\xi$ is a nuisanse parameter}. \citet{becker2021} tests this simple model with simulations assuming infinite quasar lifetimes. For cases they tested, where the true MFP $>1$ pMpc, they find that this model recovers the true MFP accurately. Recently, two independent studies \citet{satyavolu2024,roth2024} have also tested this model with different simulations. For the cases where the true MFP $\approx 1$ pMpc, they find that the model works surprisingly well even in scenarios where there are neutral patches in the IGM.

There are many uncertain factors in simulating quasar sightlines. For example, the morphology of the reionization, the distribution of LLS/DLAs and IGM neutral patches around these haloes, the lifetime of the quasars can all potentially impact the results.
It is thus valuable to further investigate this problem with different sets of simulations.
Besides, \citet{becker2021, satyavolu2024, roth2024} have only tested cases where the true MFP $\approx 1$ pMpc. Whether the formula would break down in a lower MFP scenario is not yet explored.

In this paper, we use Cosmic Reionization On Computers (CROC) simulations \citep{gnedin2014,gnedin2014b, gnedin2017} to investigate the MFP recovery procedure presented in \citet{becker2021, zhu2023}.
CROC is a suite of radiative transfer cosmological simulations with high resolution. Different from the simulations used in \citet{becker2021,satyavolu2024,roth2024}, it has fully-coupled radiative transfer with hydrodynamics, thus models the galaxy formation and the reionization processes in a more self-consistent way. In this study, we investigate three snapshots in three very different MFP regimes and examine the uncertainties in recovering the true MFP using \B21's formalism. The paper is organized as follows: in Section \ref{sec:method}, we describe the simulation details and \citet{becker2021}'s recovering procedure. In Section \ref{sec:results}, we show the synthetic profiles and fitting results. In Section \ref{sec:diss}, we explore a few variations in fitting, compare with previous studies and discuss the implications on quasar lifetime. Conclusions are presented in Section \ref{sec:concl}.

\begin{table}[]
\begin{tabular}{|p{0.3\columnwidth}|p{0.3\columnwidth}|p{0.3\columnwidth}|}
\hline
redshift & MFP [pMpc]   & MFP [pMpc] \\
 & random places &QSO-like sightlines \\ \hline
$z=6.8$      &       0.35        &        0.34             \\ \hline
$z=6.1$      &      2.1         &        1.8             \\ \hline
$z=5.4$      &      6.8         &        6.3             \\ \hline
\end{tabular}
\caption{The MFP directly measured from the simulation, with sightlines starting from random places in the box, and sightlines centered on the 20 most massive halos \citep{fan2024}.}\label{table:sightline_start}
\end{table}

\begingroup 
    \setlength{\tabcolsep}{6pt} 
    \renewcommand{\arraystretch}{1.5} 
    \begin{table*}[]
    \vspace{4mm}
\begin{tabular}{||p{35mm}|p{25mm}|p{25mm}|p{25mm}|p{25mm}|}
\hline
 MFP [pMpc] different definitions ($\pm 0.1$ pMpc) 
 & mean of  \newline profile crossings
 & crossing of \newline mean profile  & median of profile crossings & crossing of \newline median profile \\ \hline
$z=6.8$                                  & 0.4                       & 0.4                         & 0.4                      & 0.4                              \\ \hline
$z=6.1$                                  & 2.0                        & 1.8                          & 1.8                       & 1.8                                 \\ \hline
$z=5.4$                                  & 6.8                        & 5.9                          & 6.4                      & 6.4                                \\ \hline

\end{tabular}
\caption{Mean free path for different averaging methods. From left to right, we show the MFP calculated from: the mean of the crossing points from the 1000 sightlines, the crossing point of the mean profile, the median of the crossing points, and the crossing point of the median. 
 }\label{table:mfp_averaging}
\end{table*}
\endgroup

\section{Methods}
\label{sec:method}

\begin{figure*}
    \includegraphics[width=0.33\textwidth]{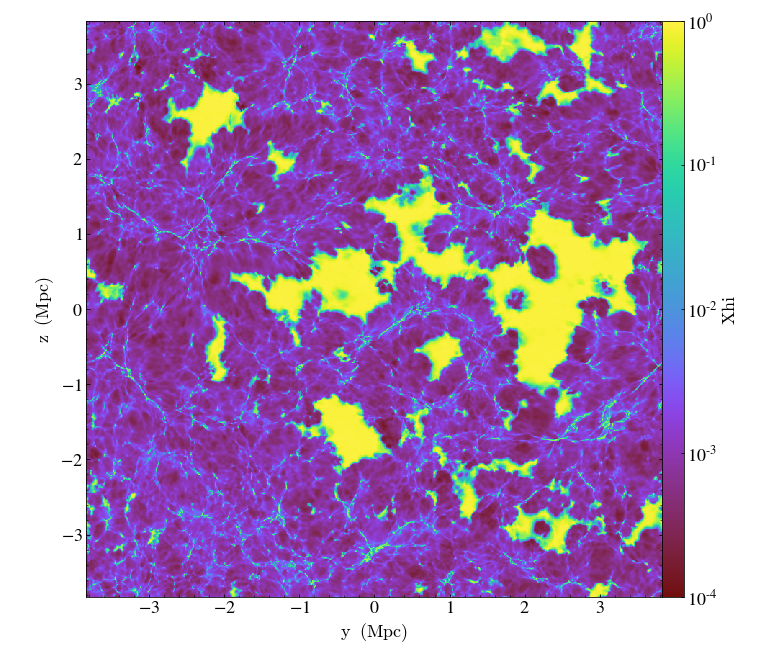}
    \includegraphics[width=0.33\textwidth]{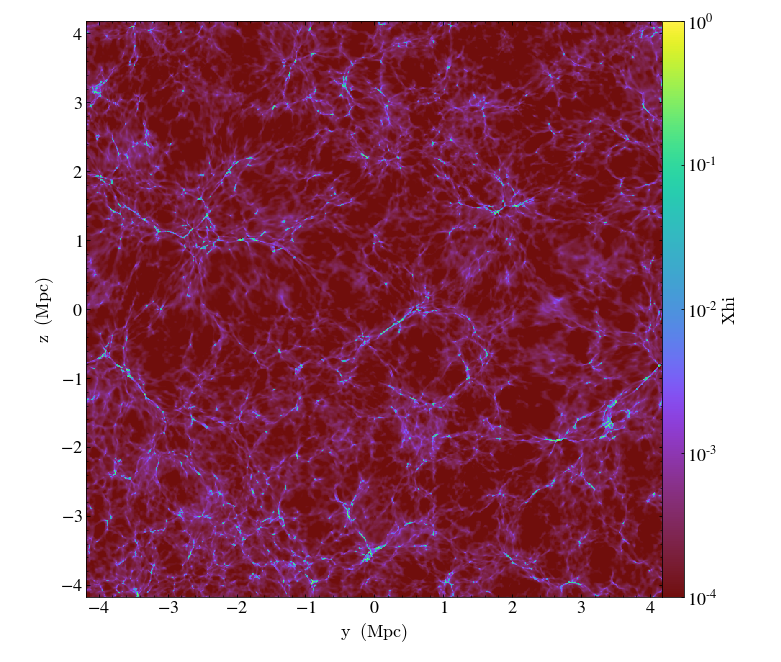}
    \includegraphics[width=0.33\textwidth]{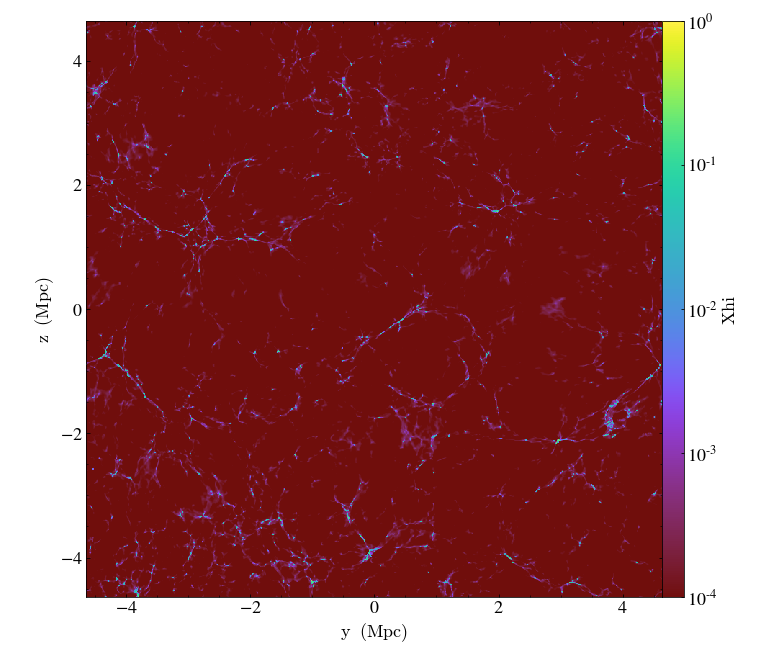}
    \caption{Neutral fraction map of thin slices (50 pkpc in depth) at different redshifts.  From left to right: $z=$6.8 when the volume-weighted neutral fraction is $\xHIv \approx0.1$, $z=$6.1 ($\xHIv \approx 2\times 10^{-4}$) and $z=5.4$ ($\xHIv \approx 8\times 10^{-5}$). Coherent neutral islands remain visible at $z=6.8$.}\label{fig:neutralfractionproj}
\end{figure*}

In this paper, we investigate the accuracy of \citet{becker2021}'s procedure in recovering the true MFP. To this end, we create sythetic quasar spectra and fit them using \citet{becker2021}'s model.

\subsection{Synthetic spectra from CROC simulations}
We make synthetic spectra of slightlines drawn from one of the CROC simulation boxes, B40F\citep{gnedin2014,gnedin2014b, gnedin2017}. This simulation box has the latest reionization history in all the six $\rm L_{box}$ = 40 $h^{-1}$cMpc CROC boxes. The properties of the IGM of this simulation are detailed in \citet{fan2024}. We focus on three representative snapshots at $z=5.4, 6.1, 6.8$. These three snapshots represent three distinctive timestamps: $z=6.1$ is when all the neutral patches in the box just disappeared ($\approx 50$ Myr ago); $z=6.8$ is when there are still neutral patches in the box; while $z=5.4$ is $\Delta t \approx 200$ Myr after all the neutral patches disappeared, a significant long time where pressure-smoothing shows visible effect \citep{cain2024}.
The distribution of neutral hydrogen in these three snapshots are very different, as shown in Figure \ref{fig:neutralfractionproj}. 

The simulation does not include individual quasars as ionizing sources. We model quasar absorption spectra by post-processing.  In each snapshot, we choose the 20 most massive halos, from which we each draw 50 sightlines isotropically.
{We calculated the mean free path defined as the average length one photon travels before it is absorbed by neutral gas. Specifically, for each sightline, we calculate the distance $d$ for which
\begin{equation}\label{eqn:def}
    \frac{1}{\left< \sigma_{\rm 912} \right>} =  \int_0^{d}n_{\rm HI} dr,
\end{equation}
where $\left< \sigma_{\rm 912}\right> =6.3\times 10^{-18} \rm\  cm^2$. Then the MFP is defined as the average of $d$ from all 1000 sightlines at each redshift. 
As the MFP measured using sightlines starting from massive halos can be different from that starting from random places, we calculate the value both starting from random locations (unbiased) and starting from the 20 most massive halos (biased).
 We list the values in Table \ref{table:sightline_start}, and refer to \citet{fan2024} for a detailed investigation of how MFP differs in different environments. 
 \citet{fan2024} has also found that the mean free path changes slightly with the exact averaging scheme. As an example, here we calculate another definition of MFP using LyC profiles.
We calculate the optical depth bluewards of LyC by convolving the transmission after each gas cell along the sightline:
\begin{equation}\label{eq:trans_sim}
    \tau (\nu) =\int_0^{\infty} n_{\rm HI} \sigma_{\rm LyC}(\nu' -\nu) dr
\end{equation}
where $r$ is the distance from the gas cell to the quasar, $\sigma_{\rm LyC}$ is the photoionization cross-section and 
$\nu'$ is the equivalent frequency corresponding to the gas cell. For each sightline, we make an absorption spectrum from the Lyman edge to $820$ \AA ~ with a resolution of $100~\rm km/s$, similar to the data resolution presented in \citet{becker2021}. We use four different methods to define a MFP using LyC profiles and list the numbers in Table \ref{table:mfp_averaging}: 1. calculate the distance between the Lyman limit to the crossing point between the LyC profile and $e^{-1}$ for each sightline, and take the average; 2. calculate the mean LyC profile by taking the average at each pixel and measure the distance between the Lyman limit to the crossing point in this mean profile; 3. similar to 1, but take the median; 4. similar to 4., but using the median profile. We find they differ slightly but within the $10\%$ level.
For convenience, in this paper, we define the ``true'' MFP as the unbiased value, measured starting from random places listed in Table \ref{table:sightline_start}.}

We post process the sightlines with a 1D radiative-transfer code described in \citep{chen2021a}. {This code solves the evolution of ionization fraction of hydrogen and helium, considering the effect of photoionization from both the quasar and the cosmic ionizing background as well as collisional ionization. Temperature evolution is also calculated self-consistently, including photoheating, recombination cooling, collisional cooling from ionization and excitation, Bremstrahlung cooling, inverse Compton cooling and cooling from Hubble expansion. Different from the code used in \citet{bolton2007,davies2016,satyavolu2024}, this code implements an adaptive prediction-correction scheme to solve each gas cell and pass the transmission to the next cell. This is motivated by the drastic different ionization timescale around quasar and
allows high temporal resolution. Both codes consider the time delay effect along the line-of-sight of the observer's frame and adopts infinite speed of light to calculate the transmission spectra. We refer interested readers to Sec. 2.2 and Fig. 3 in \citet{chen2021a} for more technical details.}

The quasar ionizing spectrum is assumed to be a power law $L_{\nu}= L_{0} {\left(\frac{\nu}{\nu_0}\right)}^{\alpha}$ with index $\alpha=-1.7$  and the total ionizing photon number rate 

\begin{equation}
  \dot{N}_{\rm tot} = \int_{\nu_{912\ang}}^{\infty} \dot{N}_{\nu} d\nu
  \label{eqn:photon_rate}
\end{equation}
is set to be $1.4\times 10^{57} \rm s^{-1}$. Using the quasar template in \citet{lusso2015}, this translates to magnitude $\Mag=-26.73$. We also run another set with a dimmer quasar $\dot{N}_{\rm tot}=3.5\times 10^{56} \ \rm s^{-1}$ ($\Mag=-25.08$), mimicking the LRIS sample in \citet{becker2021}.





\subsection{Becker+21/Zhu+23 model}

\citet{becker2021} has proposed a formula to recover the MFP without quasar radiation by fitting the observed spectra with a simplified model. 
Here we show their main formalism and refer the reader to the original paper for details. 

In \citet{becker2021}, the LyC profile is formulated as 
\begin{equation}
\begin{split}
 &   \tau(z_{912})  =\int_{r=0}^{r}\kappa (r', z_{912})dr'\\
 &= \frac{c}{H_0 \Omega_m^{1/2}}(1+z_{912})^{-\alpha} \int^{z_q}_{z_{912}}\kappa_{912}(z')(1+z')^{-\alpha-2.5}dz',
\end{split}
\end{equation}
{where $z_{912}$ is the redshift at which a photon emitted at redshift $z_q$ is redshifted to the Lyman limit,}
$\kappa (r', z_{912})$ is the opacity contributed from gas cell at distance $r'$ to that wavelength, and $\alpha$ is the photoionization cross-section slope at the Lyman edge. Note that this equation is the same as Eq. \ref{eq:trans_sim} except that analytical approximations of cosmology and cross section slope at the Lyman edge are used. \footnote{\B21 has set the cross-section index at the ionization potential edge to be $\alpha=2.75$, while in our RT1D code the index is 2.99 at Lyman edge and 3.99 at infinity. We tested changing $\alpha$ in the fitting formula $$\tau (z_{912}) = \frac{c}{H_0 \Omega_m^{1/2}}(1+z_{912})^{\alpha} \int^{z_q}_{z_{912}}\kappa_{912}(z')(1+z')^{-\alpha-2.5}dz' $$ and found that the profile dependence on this number is negligible --- the difference is less than 1\%  in the LyC when changing $\alpha$ from 2 to 4. }
The Becker's model has three parameters ($\kappa_{912}^{\rm bkg}$, $\Req$, $\xi$ ).
The first parameter, $\kappa_{912}^{\rm bkg}$ is the constant background opacity of the IGM, {which is approximately $1/\mfp$ at the redshifts of interest}. Without the radiation from the quasar, the model results in a profile where $\kappa(z_{912})=\kappa_{912}^{\rm bkg}$ at all distances.
The rest two parameters are related to quasar radiation and the response of IGM to it. The physical meaning of $\Req$ is the distance where the ionizing radiation from quasar $\Gamma_{\rm qso}$ is equal to that from the background $\gbg$, assuming only geometric dilution. Thus, $\Req$ is a parameter characterizing quasar brightness
$\Req \propto (\frac{L_{\rm qso}}{4\pi \Gamma_{\rm bkg}})^{0.5}$.
When quasar radiation is considered, $\kappa_{912}$ cannot be assumed constant but depends on the distance from the quasar. 
To approximate this,  a simple power-law relation is introduced with a parameter $\xi$:
\begin{equation}
    \kappa_{912}=\kappa_{912}^{\rm bkg} \left(1+\frac{\Gamma_{\rm qso}}{\gbg}\right)^{-\xi},
\end{equation}
where the ionization rate ratio between the quasar and the background $\frac{\Gamma_{\rm qso}}{\gbg}$ can be solved subsequently using 
\begin{equation}\label{eq:kappa_xi}
    \Gamma_{\rm qso}(r+\delta r)= \Gamma_{\rm qso} (r) \left(\frac{r+\delta r}{r}\right)^{-2}e^{-\kappa_{912}(r)\delta r}.
\end{equation}

The three parameters ($\mfp, \Req, \xi$) thus determine the transmitted LyC flux. \citet{becker2021} further finds that the three parameters impact the shape of the spectrum in different ways, thus argues that the profile could constrain $\mfp$ well even when other parameters are uncertain.

In the next section, we examine the simulated LyC profiles from different snapshots and in the presence of different {fixed} quasar lifetimes, $t_Q$. 
{
We then fit the profiles with \citet{becker2021}'s formulism with fixed nuisanse parameters $\Req$ and $\xi$ to understand the best-fit $\mfp$ dependence on $t_q$. In Section \ref{sec:diss}, we investigate more flexible fitting schemes, including simultanously fitting three parameters. As this study focus on testing the fitting scheme in a radiative transfer simulation, we do not include foreground Lyman series in either making synthetic observation or model fitting. We note that this is an important step in analyzing real observational data.}

\begin{table}[]\label{table:fitting}
\begin{tabular}{|l|l|l|l|l|l|}
\hline
$z$     & $t_Q$     & $\dot{N} $         & $\Req$  & $\mfp$  & MSE \\ 
    & Myr& $\rm ~ [s^{-1}]$   & [pMpc]  & [pMpc] &  \\ 
\hline
6.8 & 1 Myr  & $1.4\times10^{57}$ & 36              &       0.14       &  0.00038    \\ \hline
6.8 & 10 Myr & $1.4\times10^{57}$ & 36                     &    0.42          &  0.00023    \\ \hline
6.8 & 30 Myr & $1.4\times10^{57}$ & 36                     &       0.60       &   0.00010  \\ \hline
6.8 & 1 Myr  & $3.5\times10^{56}$ & 18                     &        0.15      & 0.00037   \\ \hline
6.8 & 10 Myr & $3.5\times10^{56}$ & 18                     &       0.36       & $8.3\times10^{-5}$    \\ \hline
6.8 & 30 Myr & $3.5\times10^{56}$ & 18                     &         0.5     &   $0.1\times10^{-4}$  \\ \hline
6.1 & 10 Myr & $1.4\times10^{57}$ & 16                     &      2.5        &  $5.2\times10^{-5}$   \\ \hline
6.1 & 10 Myr & $3.5\times10^{56}$ & 8                      &      2.3        &  $1.9\times10^{-5}$    \\ \hline
5.4 & 10 Myr & $1.4\times10^{57}$ & 10                     &       7.2       &  $9.4\times10^{-6}$   \\ \hline
5.4 & 10 Myr & $3.5\times10^{56}$ & 5                      &        6.8      &  $1.2\times10^{-5}$    \\ \hline
\end{tabular}
\caption{Best-fit $\mfp$ (5th column) {(with $\xi$ fixed to 0.67)} for different redshift (first column) and quasar brightness (second column). $\Req$ (4th column) is fixed according to the quasar brightness and the median ionizing background at each $z$. Note that the parameter uncertainties on $\mfp$ for all cases are tiny ($<0.01$ pMpc), smaller than the spectral resolution ($\approx 0.1$ pMpc). The corresponding mean-squared-errors (MSEs) are reported in the last column.}
\end{table}

\section{Results} \label{sec:results}

\begin{center}
    \begin{figure*}[!t]
        \centering
\includegraphics[width=0.45\textwidth]{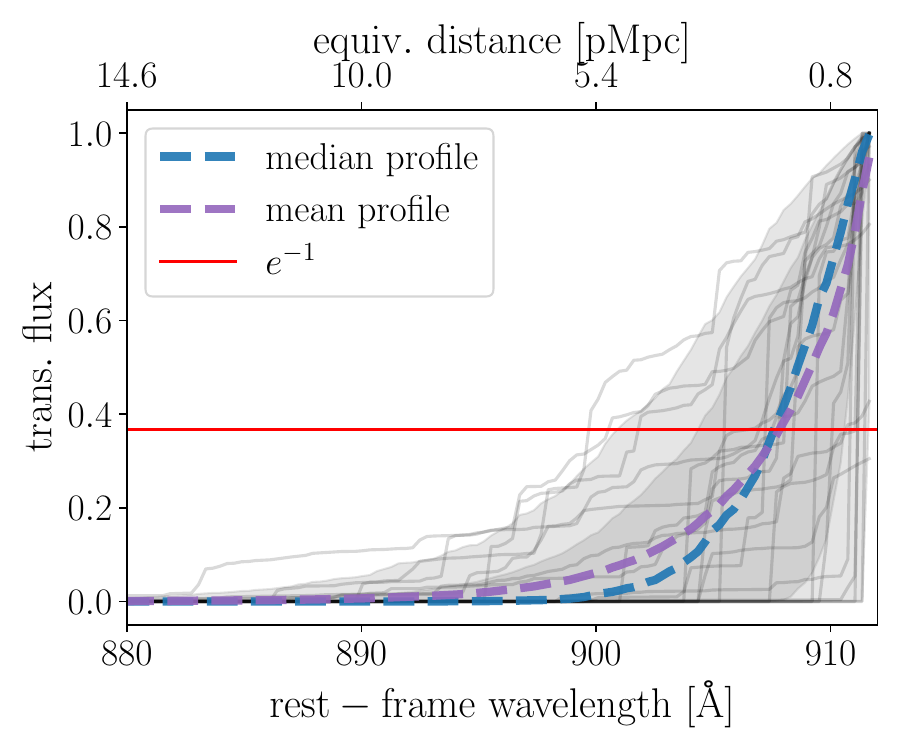}
    \includegraphics[width=0.45\textwidth]{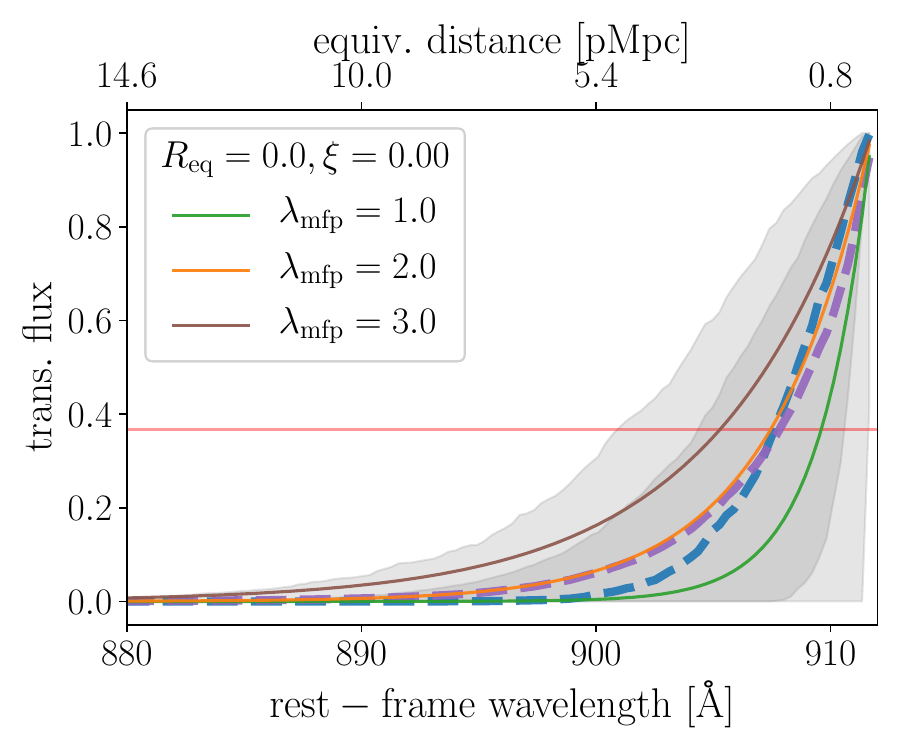}
    	\caption{Left: LyC absorption spectra for sightlines from the CROC B40F $z = 6.1$ snapshot. The blue and purple dashed lines respectively show the median and mean profiles of all 1000 sightlines. The grey shaded regions indicate the 68\% and 95\% percentile spreads. Thin grey lines show example individual spectrum. The red horizontal line show the $e^{-1}$ level. Right: same as the right with solid lines showing the models with different $\mfp$.}
        \label{fig:noq}
    \end{figure*}
\end{center}

\subsection{The Mean Free Path from LyC Profiles without Quasar Radiation}
We first examine the LyC absorption spectra and their ``averaged'' profiles, assuming there is no quasar radiation.
The left panel of Figure \ref{fig:noq} shows the profiles of LyC absorption spectra for sightlines from the CROC B40F $z=6.1$ snapshot. We illustrate example individual profiles with faint thin grey lines.  The blue and purple dashed lines respectively show the median and mean profiles of all $1000$ sightlines drawn from this snapshot.  The grey shaded regions indicate the $68\%$ and $95\%$ percentile spreads about the median.

From the individual profiles, we see that many exhibit a steep drop at $\approx 910$\AA, consistent with the excess of (sub-)LLSs (neutral hydrogen clumps with $N_{\rm HI} \approx 1.6 \times 10^{17} \rm cm^{-2}$) within the first physical Mpc \citep{fan2024}. The presence of (sub-)LLS systems in the first physical Mpc of most sightlines also drives the drop in the lower 1-$\sigma$ envelope in addition to the relative steepness of the mean profile compared with the median profile within rest-frame wavelengths above $\approx 910$\AA.  At smaller rest-frame wavelengths, corresponding to distance farther away from the quasar host, the mean profile exhibits a shallower slope than the median profile due to a small number of sightlines with high transmission that drive up the mean profile at these rest-frame wavelengths.


Often, the measured MFP corresponds to the distance to where the LyC absorption profile drops to $e^{-1}$ \citep{prochaska2009,fumagalli2013,omeara2013,worseck2014}, which we highlight in Figure \ref{fig:noq} with a red horizontal line.  The $e^{-1}$ horizontal line happens to coincide with the mean and median profiles near the intersection point of the two profiles. 
The MFP defined by the crossing point of the $e^{-1}$ threshold is $2.0$ pMpc and $1.8$ pMpc for the respective mean and median profiles. Note that both numbers have an uncertainty of $0.1$ pMpc because the spectral resolution is $100$ km/s. We summarize these measurements in the third (crossing of mean profile) and fifth (crossing of median profile) columns of Table~\ref{table:mfp_averaging} for $z=6.8$, $z=6.1$, and $z=5.4$.  We additionally report the mean and median of individual sightline MFP measurements, defined from the $e^{-1}$ crossing points measured from individual sightlines in Table~\ref{table:mfp_averaging}.

Finally, for visual reference and a direct comparison,  we show \citet{becker2021}'s models of the LyC absorption profile in the right panel of Figure~\ref{fig:noq}. 
These models assume no quasar radiation ($\Req=0$ and $\xi=0$), and we overplot models corresponding to MFP values of 1, 2, and 3 pMpc. 
If we fit the model to the mean profile by a simple 
chi-square
\footnote{We minimize the mean-squared-error in transmitted flux for all pixels from 820-912 $\rm \AA$.}, 
the best fit parameter $\mfp$ is 1.7 pMpc,
consistent with the distance to the 1/e cross point of the mean profile.
We assess the impact of quasar lifetimes and assumed analytic models on the measured MFPs in subsequent subsections.


\subsection{Dependence of Mean LyC Absorption Profiles on Quasar Lifetime}\label{sec:profile_quasarlifetime}
\begin{figure*}
    \centering
    \includegraphics[width=0.3\textwidth]{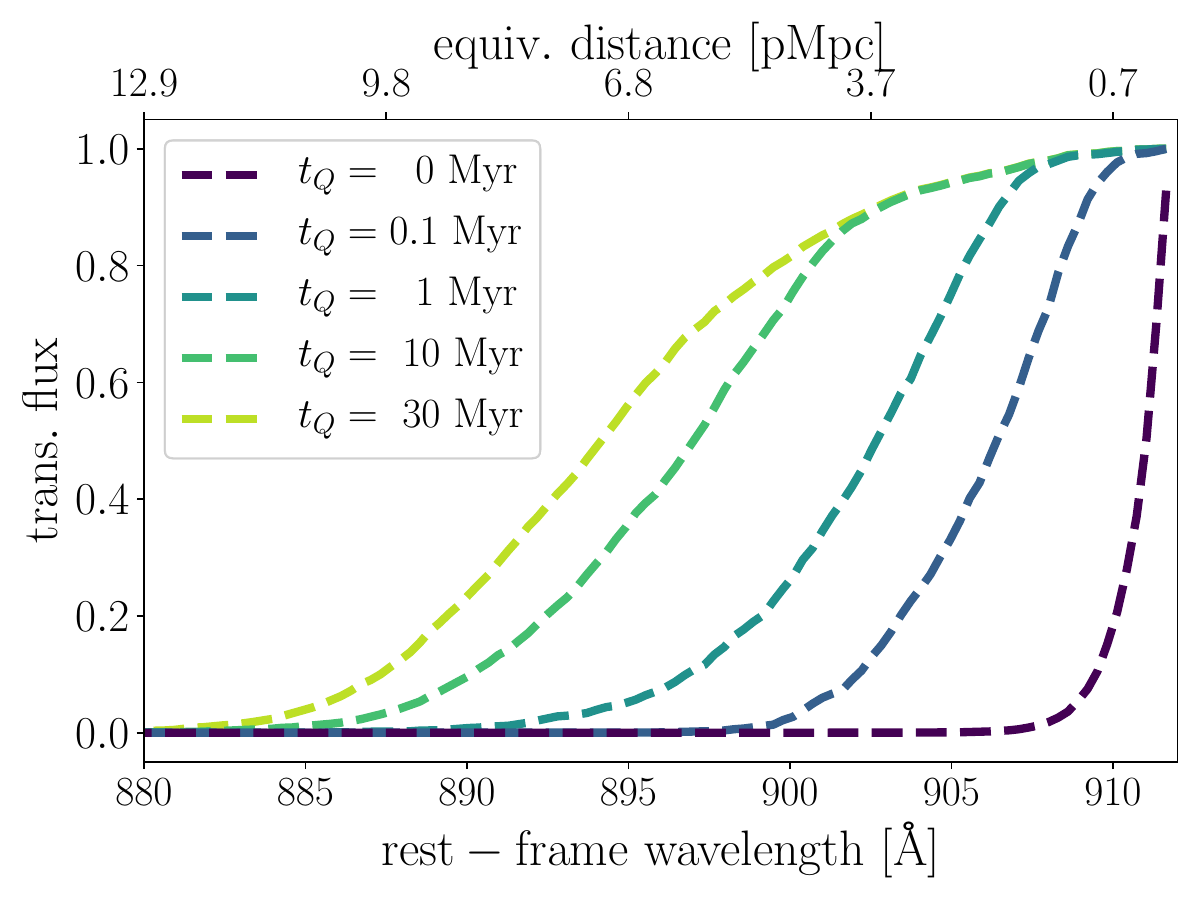}
    \includegraphics[width=0.3\textwidth]{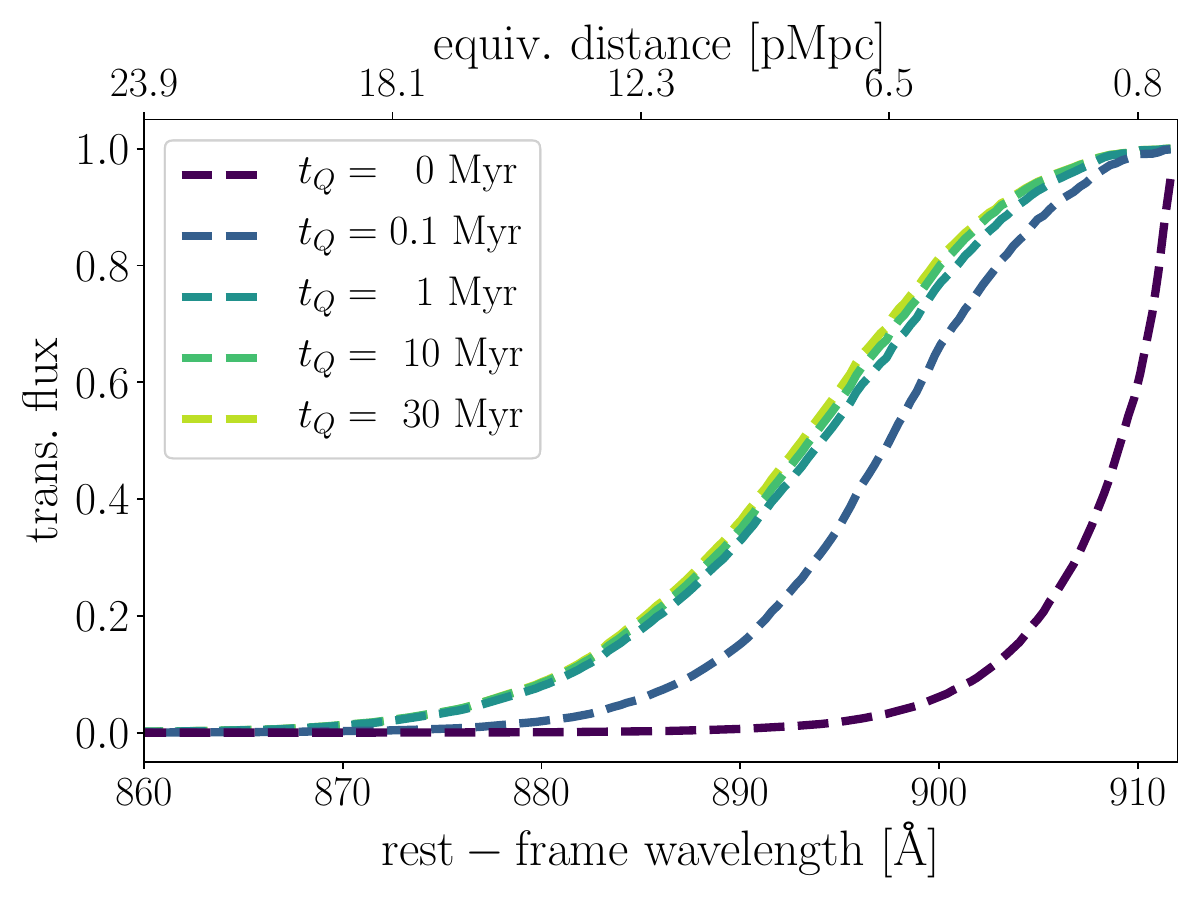}
    \includegraphics[width=0.3\textwidth]{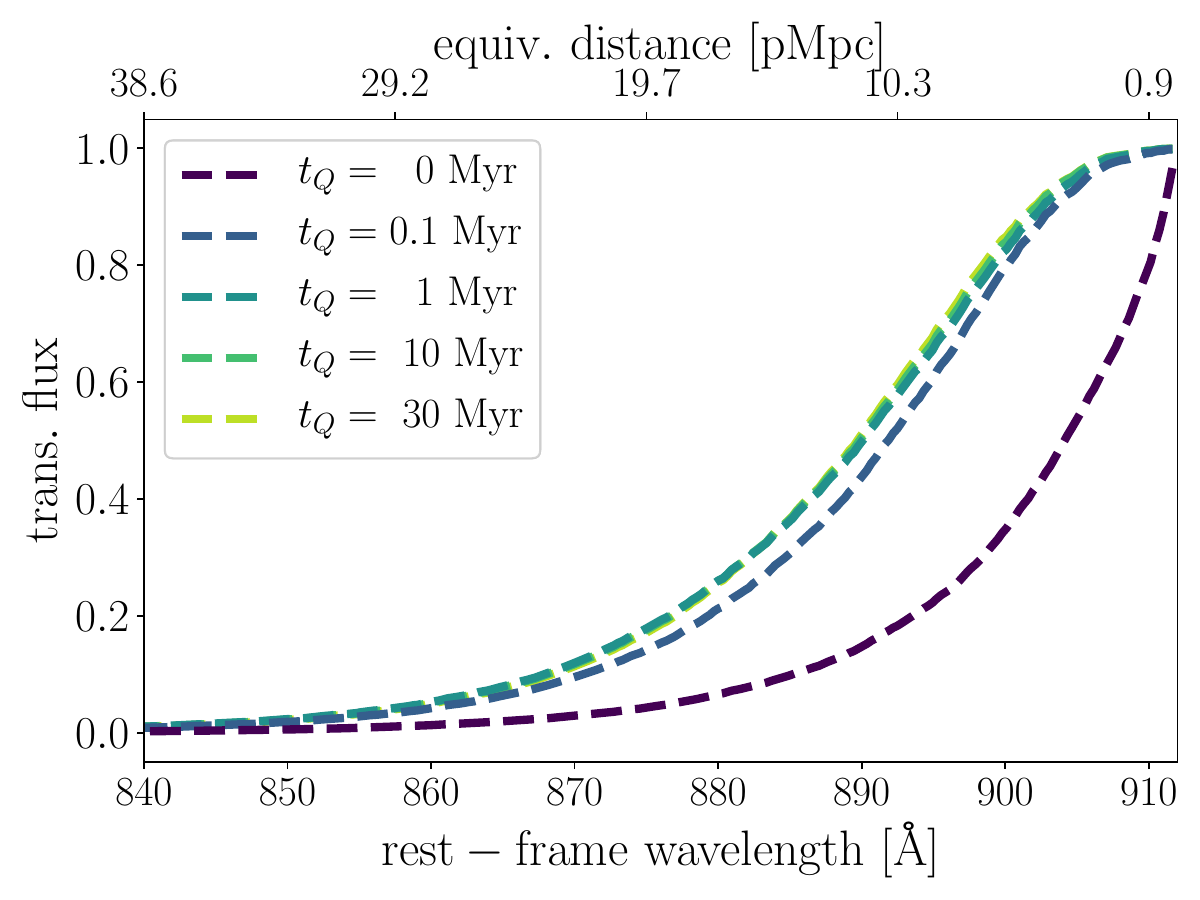}
    \includegraphics[width=0.3\textwidth]{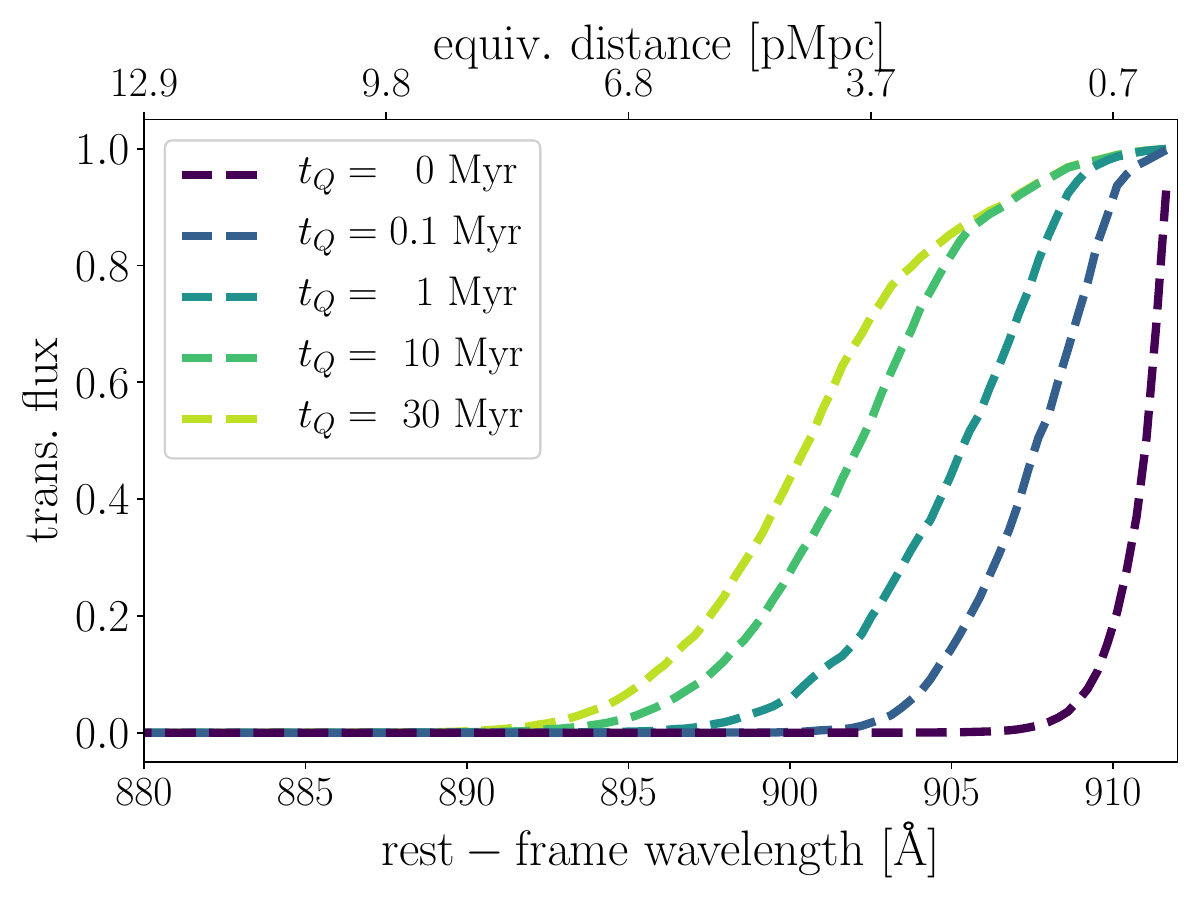}
    \includegraphics[width=0.3\textwidth]{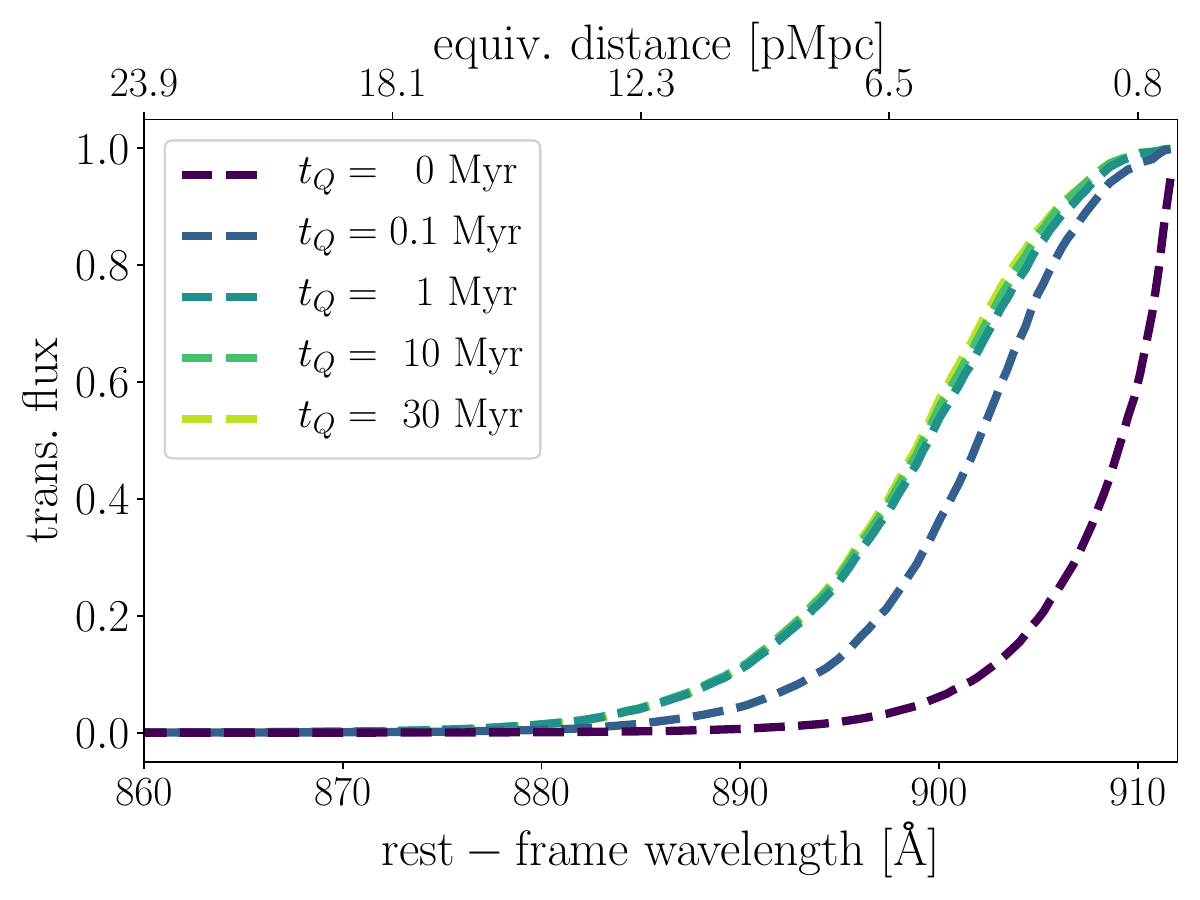}
    \includegraphics[width=0.3\textwidth]{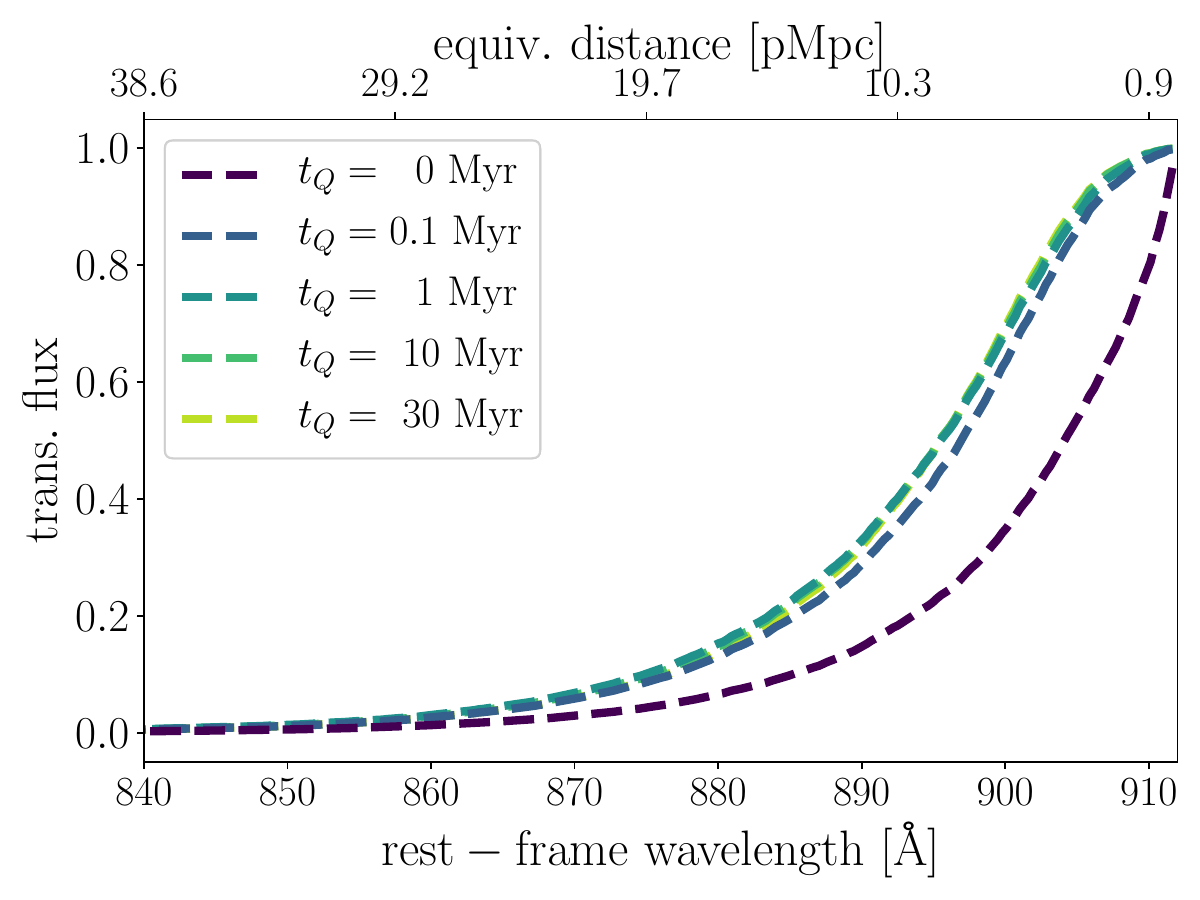}
    \caption{Mean LyC absorption profiles in the presence of a quasar. From left to right: mean profiles of 1000 sightlines at $z=$6.8, 6.1, and 5.4. Top panel: an example bright quasar case ($\dot{N}=1.4\times 10^{57} \rm~ s^{-1} $). Lower: an example faint quasar ($\dot{N}=3.5\times 10^{56} \rm~ s^{-1} $). Each line corresponds to the response of the mean profiles to varying quasar lifetime, $t_Q=$0~Myr, $0.1$~Myr, $1$~Myr, $10$~Myr and $30$~Myr.  The mean profiles exhibit the most sensitivity to quasar lifetime at $z=6.8$, when the volume-weighted neutral fraction of the box is $\sim10\%$.}
    \label{fig:profile_diff_tQ_all}
\end{figure*}

Figure~\ref{fig:profile_diff_tQ_all} shows the mean LyC absorption profile of 1000 sightlines for different quasar lifetimes.  The left, middle, and right panels each correspond to redshifts of $z=6.8, 6.1, 5.4$. The top panels show profiles resulting from the presence of a bright quasar, emitting ionizing photons with rate $\dot{N}=1.4\times 10^{57} \rm~s^{-1}$, as defined in Equation~\ref{eqn:photon_rate}.  The lower panels show profiles in the presence of a faint quasar of $\dot{N}=3.5\times 10^{56} \rm~s^{-1}$.  The general trends with quasar brightness are as expected. After the quasar turns on, as there are more photons to ionize the sightline path, the profile shows the same level of absorption on longer length scales.  For the same reason, a dimmer quasar ($\dot{N}=3.5\times 10^{56} \rm ~ s^{-1}$, lower panels) with the same lifetime has the same level of absorption on shorter length scales and overall steeper absorption profiles compared with the bright quasar case.  However, the variation in response of the absorption profile shape to quasar lifetime depends on the redshift due to the variation in neutral island and LLS content.

The leftmost profiles show the response of absorption profiles to quasar lifetime when the neutral fraction of the box is $\sim10\%$ at $z=6.8$.  Here, we see that the profile is sensitive to the quasar lifetime across all timescales plotted ($0 \sim 30 $ Myr), in both the bright and faint quasar cases.  In each case, the quasar lifetime dictates a steep profile drop-off at distinct rest-frame wavelengths that systematically shift to lower rest-frame wavelength with increasing lifetime.
On the other hand, the absorption profile dependence on quasar lifetime is not as pronounced at later redshifts, when the neutral fraction of the box is significantly lower.   

\begin{figure*}
    \centering
    \includegraphics[width=0.24\textwidth]{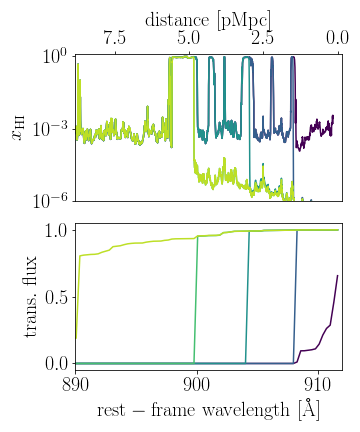}
    \includegraphics[width=0.24\textwidth]{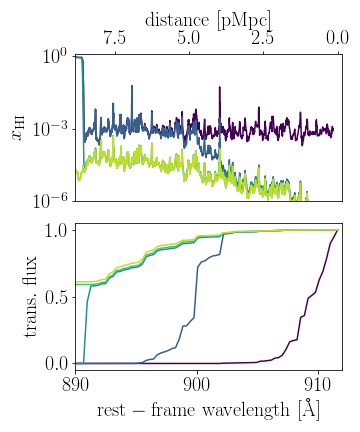}
    \includegraphics[width=0.24\textwidth]{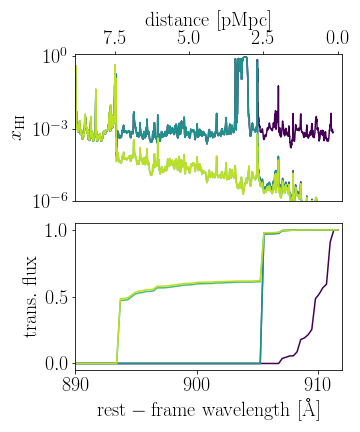}
    \includegraphics[width=0.24\textwidth]{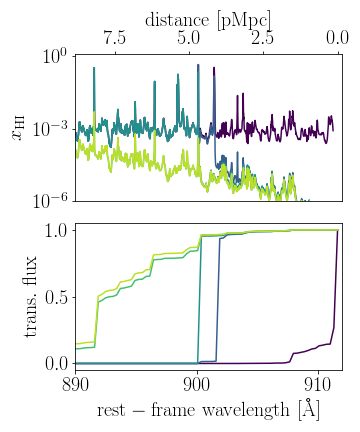}
    \caption{Neutral fraction (top) and transmitted flux (bottom) of example sightlines affected by quasar radiation on different timescales. {Line colors are the same as Fig. \ref{fig:profile_diff_tQ_all}.} Leftmost column shows an example sightline where the neutral patch contributes to a truncated flux transmission.  The second column shows a sightline that first encounters a  clump and then a neutral patch.  The third shows a sightline which encountered very dense clumps at $d\approx 2.7$ and $7.5$ pMpc and extended neutral patches at $d\approx 3$ pMpc.
    The fourth shows a case with many dense clumps along the sightline that affect the propagation of ionizing photons.}
    \label{fig:z68_examples}
\end{figure*}

To investigate the cause of pronounced dependence of the absportion profiles on quasar lifetime at $z=6.8$, we examine individual profiles. In particular, we focus on $t_Q=1$ Myr and $t_Q=10$ Myr, when the mean profiles of which minimally differ at $z=6.1$ and 5.4, but exhibit significant difference at $z=6.8$.
Visually examine the sightlines at $z=6.8$, we find that for about $\sim 50\%$ of them, the main structures that slow down the quasar I-front propagation between $1$ Myr and $10$ Myr are extended neutral patches, the other $\sim 30\%$ sightlines are to dense clumps (LLS/DLAs), and the rest $\sim 20\%$ sightlines are due to a combination of both. The drop-off feature in the mean profiles, e.g. where equivalent distance = $4.5$~pMpc in the $10$~Myr profile, likely corresponds to the typical distance in which the quasar radiation sweeps out most of the neutral gas content. Outside this distance, the quasar radiation has not yet ionized the neutral patches or LLS/DLAs that contribute to the drop-off in profiles at $z=6.8$.  We show four randomly selected sightlines from the $z=6.8$ snapshot in Figure~\ref{fig:z68_examples}, which exemplify the diversity of sightlines at this redshift.  Each line color corresponds to the same quasar lifetime as shown in the legend of Figure~\ref{fig:profile_diff_tQ_all}. Among them, we can find that the evolution of the LyC profile is neutral patch dominated, LLS dominated or a combination of both.

The leftmost panel shows a sightline that goes through many neutral patches with various sizes, all of which slow down the propagation of quasar I-front. Some neutral patches along this sightline are very wide ($>0.5$ pMpc, e.g., the ones at $d\approx3$ and $\approx 5$ pMpc), while others are very narrow.  Note, none of the narrow structures are dense clumps, as evidenced by the fact that these structures do not create a significant LyC absorption in the presence of a quasar with lifetime $t_Q>10$~Myr.
The second panel shows a sightline that only encounters a single extended neutral patch very far away ($d=8$ pMpc), and one slightly dense clump at $d\approx4$~pMpc. The clump at $d\approx4$ pMpc becomes optically thin within 0.1~Myr.  The extended neutral patch at $d\approx 8$~pMpc leads to different profiles between quasar lifetimes of $t_Q=1$~Myr and $t_Q=10$~Myr. 
The third panel and the fourth panel show sightlines where LLS/DLAs play a crucial role in the evolution on the change of LyC profile on $\sim 10^6$~yr scales. In the third panel, a dense structure at $d\approx 2.7$~pMpc pauses the quasar ionization front (I-front) for $\sim1$~Myr.  Even after reaching the new photo-ionization equilibrium, the dense structure still blocks almost half of the ionizing photons, as demonstrated by the drop in the LyC profile corresponding to $t_Q=10$~Myr. After the I-front passes this structure and the neutral patches immediately after it (at $d\approx 3$~pMpc), another dense structure at $d\approx 7.5$ pMpc blocks the ionizing radiation again. The fourth panel showcases a sightline that travels through many dense clumps. The ones at $\approx$4, 4.5, and 8~pMpc respectively cause the change in LyC profile on timescales of $0.1$, $1$, and $10$~Myr.

Note, at later redshifts ($z=6.1$ and $5.4$), extended neutral patches are no longer present.
 Contrary to the case at $z=6.8$ pMpc, the LyC profiles at $z=6.1$ and $z=5.4$ do not exhibit a further response to quasars after $t_Q=1$~Myr. At $z=6.1$, the background radiation, $\Gamma_{\rm bkg}$ is lower than at $z=5.4$ and there are more structures that are optically thick at all distances (see also  \citet{fan2024}). This is consistent with what we see here: at $z=6.1$, there is a visible response of the LyC profile to the quasar, even on a timescales of $\sim10^5$~yr. On the other hand, structures are relatively more transparent at $z=5.4$, allowing the propagation of quasar radiation on shorter timescales more easily than at $z=6.1$. 

In summary, we find that the extended ($\gtrsim 0.5$ pMpc) neutral patches are not the only mechanism that slows down the propagation of quasar radiation to lead to dependence of LyC profiles on quasar lifetime at $z=6.8$.  For many of the sightlines, small-scale ($<0.1$ pMpc) clumps are responsible for truncating flux transmission along the sightline, leading to a dependence of the LyC profile on quasar lifetime.
The investigation of the mean profile evolution and its interconnected dependence on the quasar lifetime and environment indicate potential issues that may arise when fitting the spectra with a simple model, such as that presented in \citet{becker2021}, for MFP measurements. In the next section, we perform the fit to synthetic spectra and quantify the bias and uncertainties on the inferred MFP.

\subsection{Model Fitting with Single Variable at Specific $t_Q$}  \label{sec:results:fit}

\subsubsection{Bias}
In this section, we perform a mean-square-error fit of the \B21 model to the synthetic spectra and test for biases in the inferred mean free path. Here, we limit our tests to cases with quasar lifetimes longer than $1$~Myr, a typical timescale indicated in \citet{morey2021}. 
To perform the fit, we fix $\Req$ to be the exact distance where the ionizing rate from unattenuated quasar radiation equals that from the background radiation\footnote{We use the median of the background radiation in the box, see \citet{fan2024}.} (see Table \ref{table:fitting}). {In this subsection, we also fix the nuisance parameter $\xi=0.67$ \citep{furlanetto2005,mcquinn2011}.}

\begin{figure*}
    \includegraphics[width=0.45\textwidth]{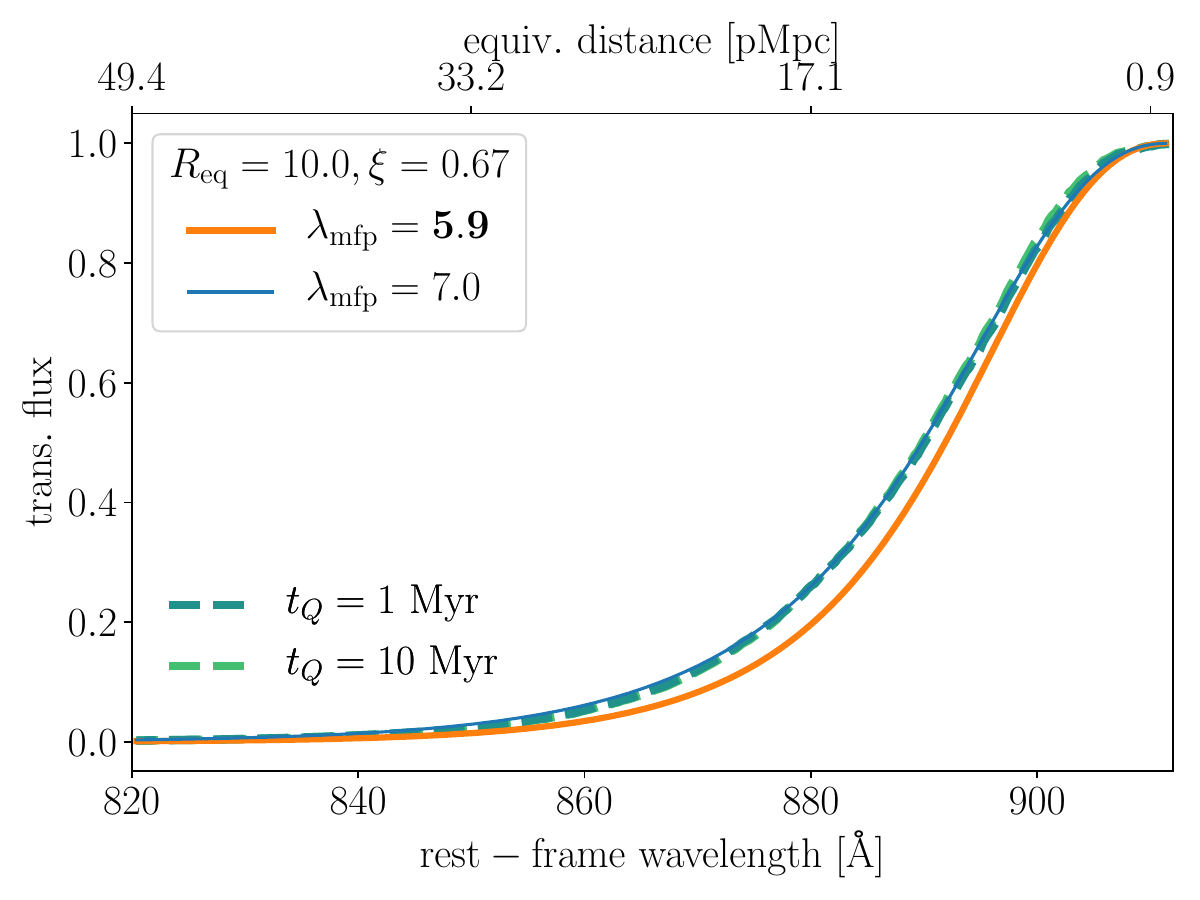}
    \includegraphics[width=0.45\textwidth]{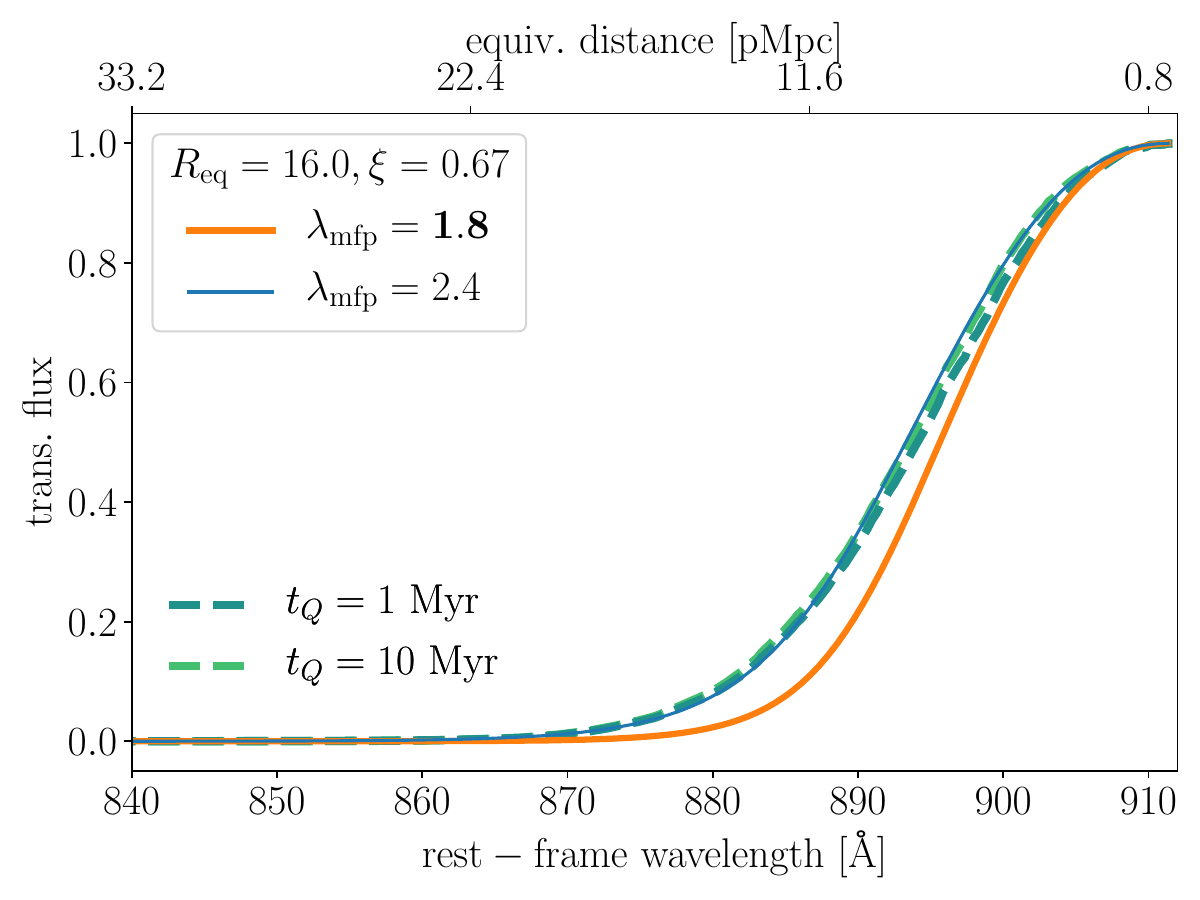}
    \caption{Mean LyC absorption profiles of 1000 sightlines at $z=5.4$ (left) and $z=6.1$ (right).  Dashed thick lines correspond to sightlines in the presence of two example quasar lifetimes, the model profile with $\mfp=$ true MFP [pMpc] overplotted in the solid thick orange curve. Resulting best-fit model profile in thin blue, labeled with the best-fit $\mfp$, which biases slightly high compared to truth.}\label{fig:fit_afterankle}
\end{figure*}

In the left panel of Figure~\ref{fig:fit_afterankle}, we show the mean LyC profiles over 1000 sightlines from $z=5.4$ for quasar lifetimes $t_Q=1$~Myr and $10$~Myr in thick dashed lines.  The profiles are near identical at this redshift.  We overplot the corresponding model profile, defined in \B21, with parameter $\mfp$ equal to the true MFP, $5.9~\rm pMpc$ (thick orange line).  We show the best-fit profile in the thin blue line, labeled with the best-fit $\mfp=7.0~\rm pMpc$ for $t_Q=1$~Myr.  The best-fit $\mfp$ to the average profile is biased high by $\approx 20\%$ compared to the true MFP directly measured from the simulation.  The right panel shows the same measurements and fits for sightlines from $z=6.1$. We see the best-fit $\mfp$ ($2.4 \rm~ pMpc$) to the average profile is biased high by $30\%$ compared to the true MFP ($1.8 \rm~ pMpc$). From this exercise, we conclude that in the regime where the true MFP is relatively large ($\gtrsim 2 $ pMpc), the MFP inferred from the best-fit model profile tends to bias slightly high when we use the correct $\Req$. As the true MFP decreases, the relative bias slightly increases.

\begin{figure}
    \centering
    \includegraphics[width=0.45\textwidth]{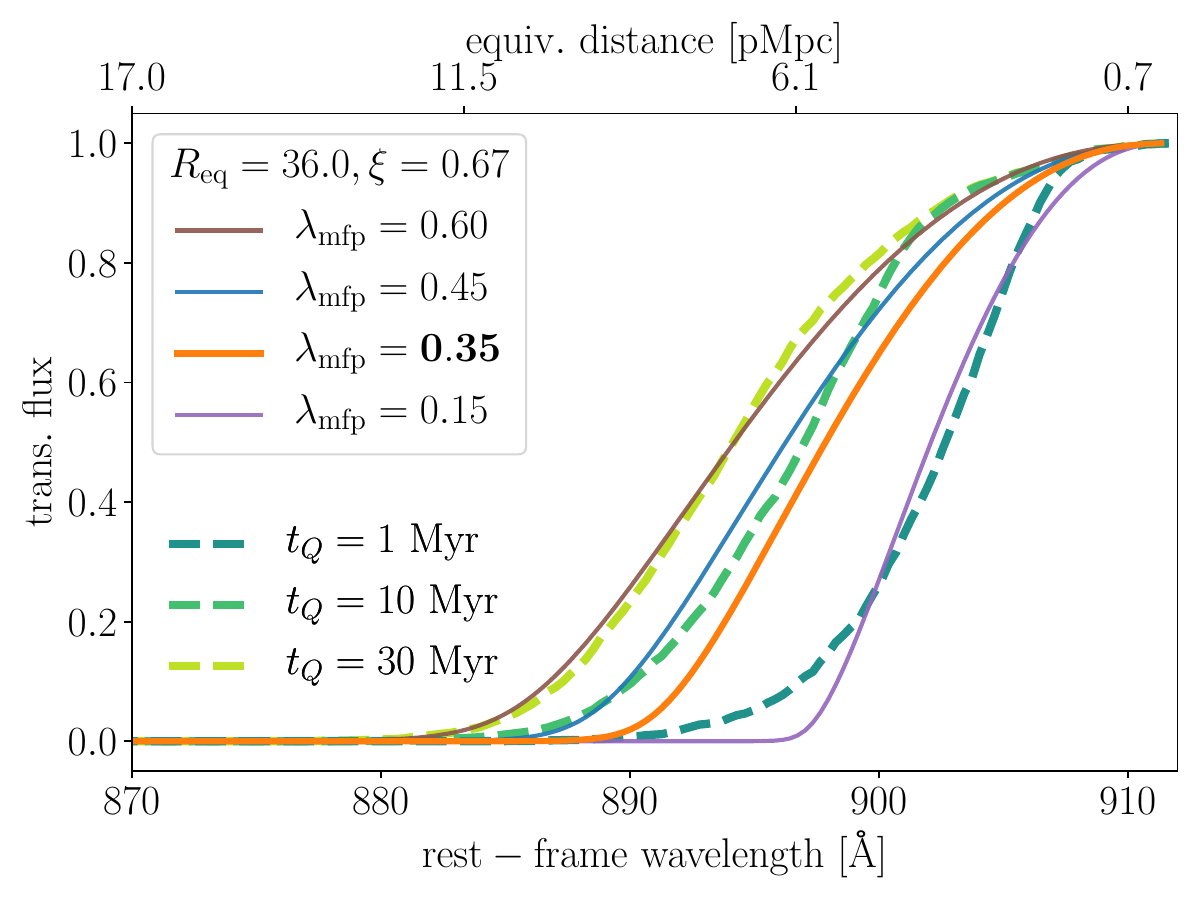}
    \caption{Thick dashed lines: mean LyC absorption profiles of 1000 sightlines at $z=6.8$ for different quasar lifetimes . Thin solid lines: model profiles with different $\mfp$ [pMpc]. Orange thick solid line: model profiles with $\mfp=$ the true MFP  $0.35$ pMpc.}
    \label{fig:fit_beforeankle}
\end{figure}

In Section~\ref{sec:profile_quasarlifetime}, we have shown that sightlines from $z=6.8$, where there are neutral islands and abundant LLS/DLA systems, have LyC profiles that strongly depend on quasar lifetimes, including timescales longer than $1$~Myr.  This regime corresponds to an epoch when the MFP is very short (true MFP$\lesssim$ 0.5 pMpc). In this snapshot, we fit the mean LyC profile for sightlines in the presence of a quasar at three different timescales: $t_Q=1,$ $10$, and $30$~Myr.  We show the mean profiles for these cases in Figure~\ref{fig:fit_beforeankle} in thick dashed lines, and the best-fit model profile in the overplotted thin solid lines.  First, we find that the model profile from \B21 no longer describes the mean profile as well as the previous cases at $z=6.1$ and $z=5.4$.  The deviation between the best-fit $\mfp$ and the true MFP is particularly large when the quasar lifetime is smaller ($t_Q\lesssim 10$~Myr). The best-fit $\mfp$ for $t_Q=1$, $10$ and $30$~Myr profiles are respectively $0.14$, $0.42$, and $0.60$~pMpc, with corresponding MSE=$3.8\times 10^{-4}$, $2.3\times 10^{-4}$ and $1.0\times 10^{-4}$.
Depending on the quasar lifetime, the best-fit $\lambda_{\rm mfp,fit}$ can be mis-estimated by a factor of few compared to the true MFP. For example, the best-fit model profile to the average profile in the presence of a $t_Q=1$~Myr quasar underestimates the true MFP by $60\%$ while the best-fit model profile to the average profile in the presence of a $t_Q=30$~Myr curve overestimates the true average MFP by $70\%$.\footnote{This significant over-estimate can be understood as that the ``true'' $\gbg$ is in fact even smaller due to the neutral patches, thus we underestimated $\Req$, which needs to be compensated by much higher $\mfp$.} This suggests that the LyC profile is probably not a good observable to measure the general MFP when the MFP is as small as $0.4$~pMpc. Instead, in this case, the LyC profile could be a powerful tool to constrain the quasar lifetime, $t_Q$. We will discuss this potential approach in Section~\ref{sec:quasarlifetime}.

\subsubsection{Sample Variance} \label{sec:samplevar}
Current facilities enable observations of $\sim 10$ quasar spectra at $z\sim 6$  for LyC absorption profile analysis. In this subsection, we explore the statistical uncertainties in $\lambda_{\rm mfp}$ due to sample variance with bootstrap. In each redshift snapshot considered from our simulations, we randomly draw 10 sightlines from the total 1000 with replacement, fit \B21 model to the mean profile of the 10 sightlines to infer a MFP, then repeat this procedure 100 times to estimate the distribution of the inferred $\lambda_{\rm mfp}$. 

\begin{figure*}
    \centering
    \includegraphics[width=0.45\textwidth]{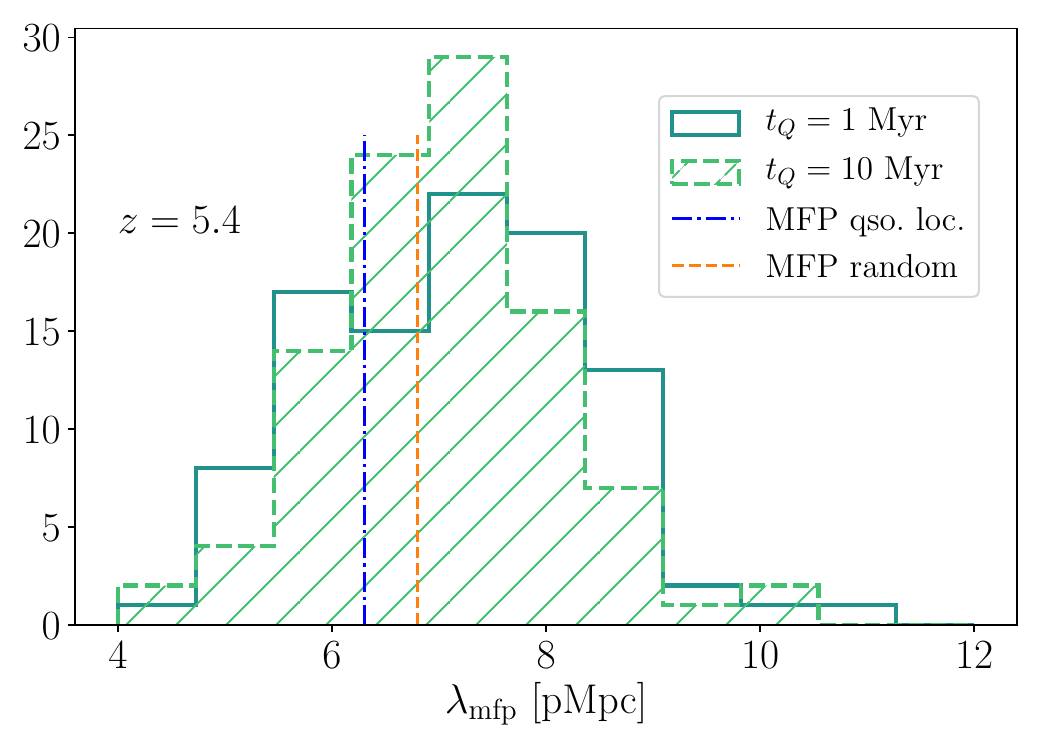}
    \includegraphics[width=0.45\textwidth]{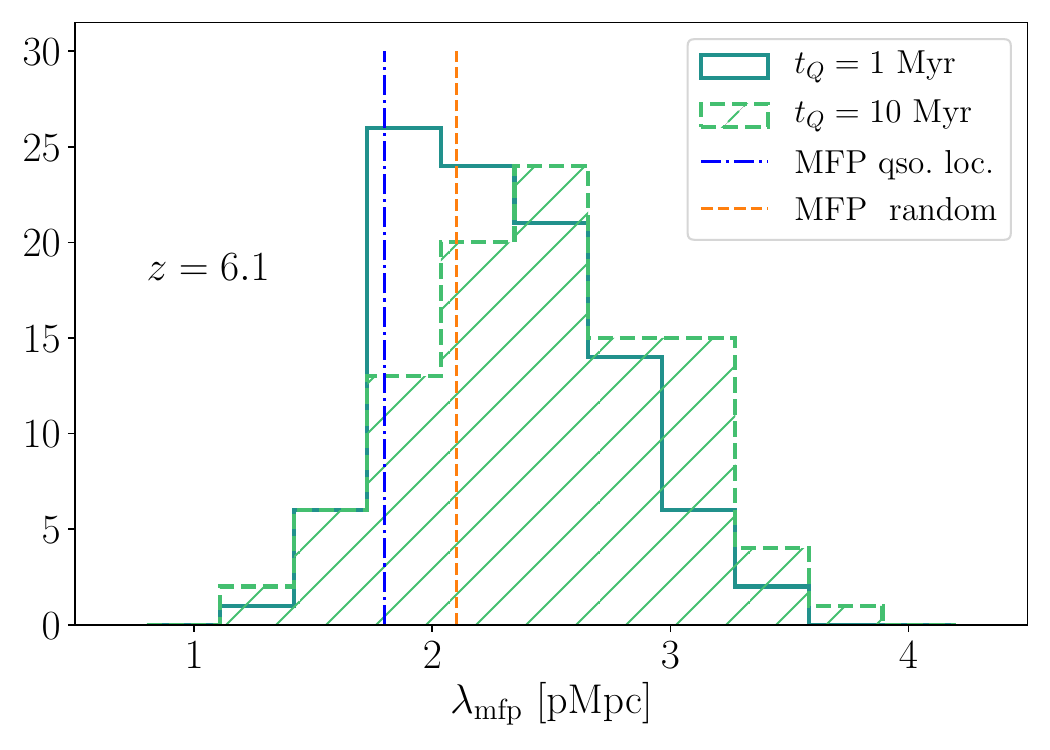}
    \caption{Sample variance of the inferred MFP from the mean LyC profile in samples of 10 sightlines at $z=5.4$ (left) and $z=6.1$ (right).  {Histograms show distributions of the best-fit $\mfp$ to the mean profile of 10 sightlines randomly drawn from the total 1000, for cases $t_Q=1$ Myr and $t_Q=10$ Myr.  Vertical blue and orange lines show the MFP directly measured from the simulation snapshot, in quasar environments (assuming quasars never turn on) and random environments, respectively.}}
    \label{fig:bootstrap_results_z54_z61}
\end{figure*}

Figure~\ref{fig:bootstrap_results_z54_z61} shows this distribution from the $z=5.4$ and $z=6.1$ snapshots.  The {solid empty and dashed shaded histograms} respectively correspond to sightlines in the presence of a quasar of lifetime $t_Q=$1~Myr and 10~Myr.  The vertical lines respectively indicate the true MFP measured in \citet{fan2024}: the dashed blue line represents the MFP measured from biased quasar-host environments while the dashed orange line represents the MFP centered on random places in the box.  We find that all four histograms have a similar relative standard deviation of $\approx 15-20\%$, with respect to the best-fit $\mfp$       inferred from the mean profiles of all 1000 sightlines. This level of spread in the best-fit MFP is similar to the true MFP spread we found in \citet{fan2024}.

We show the same distributions of the inferred $\mfp$ at $z=6.8$, including a {color-filled histogram for the distribution of MFPs for the case when $t_Q=30$~Myr.}  We find that the relative standard deviation of the $\mfp$ of sightlines seems to decrease slightly with quasar lifetime, which are respectively $\approx 30 \%$, $\approx 20 \%$ and $\approx 15 \%$ for $t_Q=1, 10, 30$~Myr.

\begin{figure}
    \centering
    \includegraphics[width=0.45\textwidth]{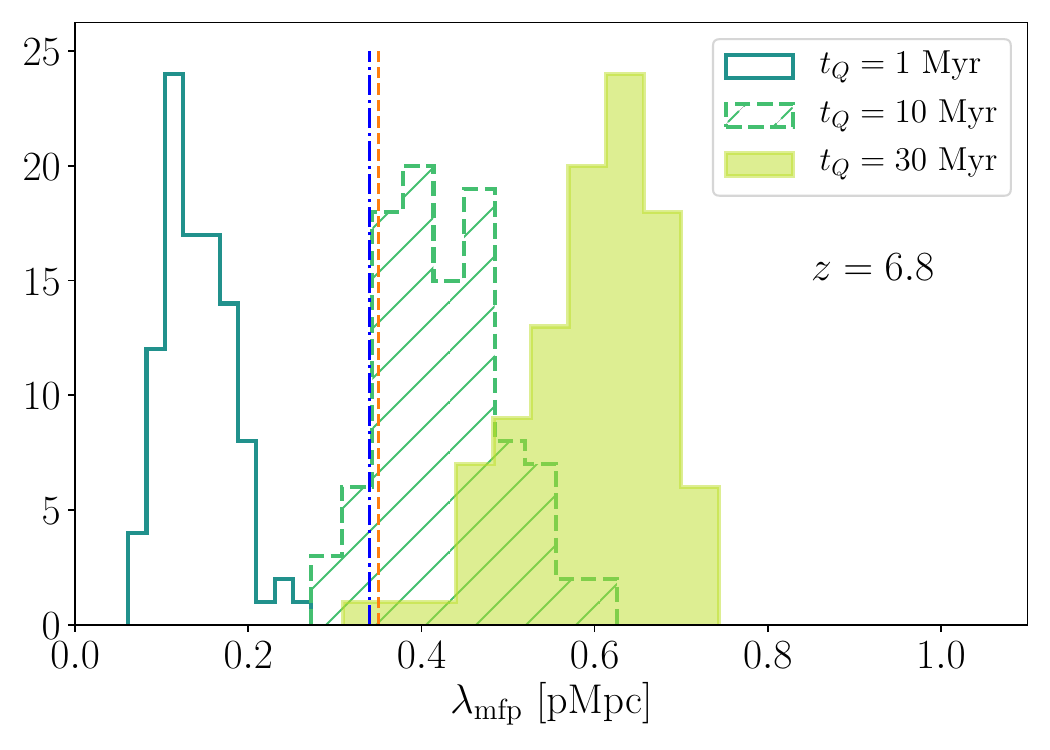}
    \caption{Same as Figure \ref{fig:bootstrap_results_z54_z61} but at $z=6.8$, with one more histogram for $t_Q=30$ Myr.}
    \label{fig:bootstrap_results_z68}
\end{figure}



\section{Discussion}
\label{sec:diss}

Here, we discuss how uncertainties on nuisance parameters affect our analysis of the mean free path. We also compare our work with similar recent studies of the mean free path in other simulations.

\subsection{Uncertainties on Nuisance Parameters}

The above analysis assumes precise knowledge of the ionizing luminosity of the quasar and the ionizing background, which is not the case in observations. 
Due to uncertainties in the shape of the quasar spectra, there could be an uncertainty of $\gtrsim 20\%$ in ionizing flux \citep{lusso2015}. An error in $\Req$ inevitably impacts the accuracy of the inferred $\mfp$. Let us consider the $z=5.4$ snapshot with $t_Q=10$ Myr as an example. If we overestimate (underestimate) the ionizing flux by $20\%$, which equivalently biases $\Req$ high (low) by $10\%$, the best-fit of $\mfp$ changes from $7.2$~pMpc to $6.8$~pMpc ($7.6$~pMpc) with corresponding MSE=$6\times 10^{-5}$ (0.0002) compared with an MSE=$3\times 10^{-5}$ that results from an unbiased $\Req$. The response of IGM opacity to quasar radiation, parametrized in $\xi$, is also uncertain. One way to mitigate the potential bias in $\mfp$ due to these nuisance parameters is to free the parameter $\Req$ and $\xi$ completely, and let the data determine all three parameters simultaneously. We explore this in Sec. \ref{sec:fit_3_par}.

Note, $\Req$ explicitly depends on $\Gamma_{\rm bkg}$, which is a highly uncertain parameter closely related to $\lambda_{\rm mfp}$. In fact, in some models, $\Gamma_{\rm bkg}$ is determined by the MFP \citep{meiksin2004,mcdonald2005,wyithe2006,mesinger2009,mcquinn2011,davies2016}. We perform a fitting test with correlated $\mfp$ and $\Req$ in order to gain insights in parameter dependency in \ref{sec:tie_mfp_Req}.

\subsubsection{Simultaneous Parameter Fitting: $\mfp$, $\Req$ and $\xi$}
\label{sec:fit_3_par}
In this section, we free all three parameters and perform the fit in our bright quasar case ($\dot{N}=1.4\times10^{57} ~\rm~ s^{-1}$). We place boundary constraints on $\mfp$ and $\Req$ to be positive and on $\xi$ to be between 0 and 1. We first examine the $z=5.4$ and $z=6.1$ snapshots, where the LyC profile stablizes after $t_Q=1$~Myr.  We test the $t_Q=10$~Myr case and compare the results to Sec. \ref{sec:results:fit} where we only free one parameter $\mfp$. For $z=5.4$, we find the best fit parameters { $(\mfp \rm{[pMpc]}, \Req \rm{[pMpc]}, \xi)=(7.1, 9.6, 0.71)$} and the MSE=$7.0\times 10^{-6}$. For $z=6.1$, we find the best fit parameters $(\mfp \rm{[pMpc]}, \Req \rm{[pMpc]}, \xi)=(2.9, 13, 0.67)$ and the MSE=$7.5\times 10^{-6}$. In both cases, $\xi$ is very close to the fiducial value of $0.67$ used in \citet{becker2021,zhu2023} and $\Req$ is also close to the correct $\Req$ measured in the simulation. Not surprisingly, the constraint on $\mfp$ is also very close to the case where we fix $\xi=0.67$ and $\Req$ to be the correct value (Sec. \ref{sec:results:fit} and Table \ref{table:fitting}). From this exercise, we have shown that the data is capable of constraining all three parameters accurately in the regime where the true MFP is large ($\gtrsim 2$~pMpc).

For the $z=6.8$ case, the model in \citet{becker2021} is likely not a good description because the LyC has not yet stablized despite the quasar being on for 1~Myr.  We therefore expect that freeing all three parameters might result in unphysical results. Indeed, if we free all three parameters, the best-fit parameters are $(\mfp \rm{[pMpc]}, \Req \rm{[pMpc]}, \xi)=(1.0,10,1)$ with MSE=$5\times 10^{-5}$ for $t_Q=10$~Myr and $(\mfp \rm{[pMpc]}, \Req \rm{[pMpc]}, \xi)=(1.2,4,1)$ with MSE=$3\times 10^{-5}$ for $t_Q=1$~Myr.  Here, we find that the inferred $\Req$ is smaller than the true $\Req$ by an order of magnitude. The MFP parameter $\mfp$  also significantly deviates from truth to allow the overall shape of the model LyC to fit the spectrum. This exercise cautions us that we should not blindly fit all LyC spectra with the model from \citet{becker2021} in order to obtain a MFP measurement. In particular, the best-fit parameters can be very unphysical in the case where the true MFP is extremely small ($\lesssim 0.5$ pMpc).


\begin{table*}[]
\begin{center}
\begin{tabular}{|p{0.45\columnwidth}|p{0.2\columnwidth}|p{0.2\columnwidth}|p{0.2\columnwidth}|p{0.2\columnwidth}|}
\hline
Mean profile with mixed $t_Q$& $\mfp$& $\Req$  &$\xi$ &MSE\\
 & [pMpc]&[pMpc]& &\\ \hline
0.1 and 1 Myr&       1.2&        4.8& 0.50&$6.9\times 10^{-6}$\\ \hline
10 and 30 Myr&      1.0&        15& 0.83&$3.0\times 10^{-5}$\\ \hline
All&      0.066&        13302& 0.26&$2.6\times 10^{-5}$\\ \hline
\end{tabular}
\caption{The best-fit parameters and MSE to the mean profiles of mixed quasar lifetime $t_Q$. Here we list the result when we use unbounded upper limit for both $\mfp$ and $\Req$. Note that there is a degeneracy between $\mfp$ and $\Req$. See Section 4.2 for details. \label{tab:mix_tQ_fit}}
\end{center}
\end{table*}

\subsubsection{Fitting Test with a Power-law Relation between $\mfp$ and $\gbg$}
\label{sec:tie_mfp_Req}
In the model from \citet{becker2021}, the parameter $\Req$ explicitly depends on $\gbg$, which is closely related to $\mfp$. Therefore, it is perhaps more sensible to choose a prior with correlated $\mfp$ and $\Req$ instead of flat priors of both. 
Tying these two parameters could offer some insights into how the best-fit $\mfp$ and $\xi$ relates to each other.
For a proof of concept, here we use a power-law relation between $\log\mfp$ and $\log\gbg$ 
\begin{equation}
    \log\mfp = k \log\gbg +C
\end{equation}
to explicitly express $\Req$ in $\mfp$ when fitting. In the CROC box, B40F, we find the slope to be $k\approx 1.3$ \citep{fan2024}. Motivated by this, we eliminate free parameter $\Req$ to be $\Req=R_{\rm eq, true}\left(\frac{\mfp}{\lambda_{\rm mfp, true}}\right)^{-0.38}$.

We first fit the model with two free parameters ($\mfp$ [pMpc], $\xi$) to the $z=5.4$ snapshot. Again, we find the best-fit by minimizing $\chi^2$ for wavelength from $820-912$ \AA. For all four cases $t_Q=0.1, 1, 10, 30$ Myr, there exists a weak anti-correlation between the two parameters at the order of ${\rm Cov}(\mfp [{\rm pMpc}], \xi) \sim 10^{-5}$.
We find the best-fit to be (6.8, 0.59), (7.4, 0.68), (7.2, 0.73), (7.0, 0.77) for $t_Q=0.1, 1, 10, 30$ Myr, respectively. The longer $t_Q$ cases favor a slightly larger $\xi$ solution. In the $t_Q=0.1$ Myr case, the (sub-)LLSs have not been fully ionized, leading to a shape that is noticeably different from the other cases. The best-fit $\mfp$ is smaller than all other cases. For all other cases, with $t_Q>1$~Myr, there only exists a tiny difference in the inner region. These longer $t_Q$ cases favor a shorter $\mfp$ solution. This trend is even more obvious for the other two snapshots where the true MFP is smaller. For the $z=6.1$ snapshot, the best-fit parameters for $t_Q=0.1, 1, 10, 30$ Myr are (2.0, 0.54), (3.1, 0.61), (2.9, 0.67) and (2.8, 0.70), {while the MSE for each best-fit are $1.5\times 10^{-5}$, $8.7\times10^{-6}$, $7.5\times 10^{-6}$,  and $6.0\times 10^{-6}$, respectively. This indicates that as more (sub-)LLSs are ionized in the proximity zones, the model describes the average profile better.}

For the $z=6.8$ snapshot, the best-fit parameters for $t_Q=0.1, 1, 10, 30$ Myr are ($\mfp$ [pMpc], $\xi$)=(0.24, 0.36), (0.24, 0.53), (0.34, 0.72), and (0.55, 0.76), respectively. However, the functional shape of the model is noticeably different from the profile at $t_Q=1, 10$ Myr: the profile has a sharp drop while the model is always smooth. This results in a poor fit with MSE$>10^{-4}$ for $t_Q=1, 10$ Myr
. This indicates that in the case where there are still neutral patches around the quasar, the model is not as good as previous cases, even for systems with relatively long quasar lifetimes.

\begin{figure}
    \centering   \includegraphics[width=\linewidth]{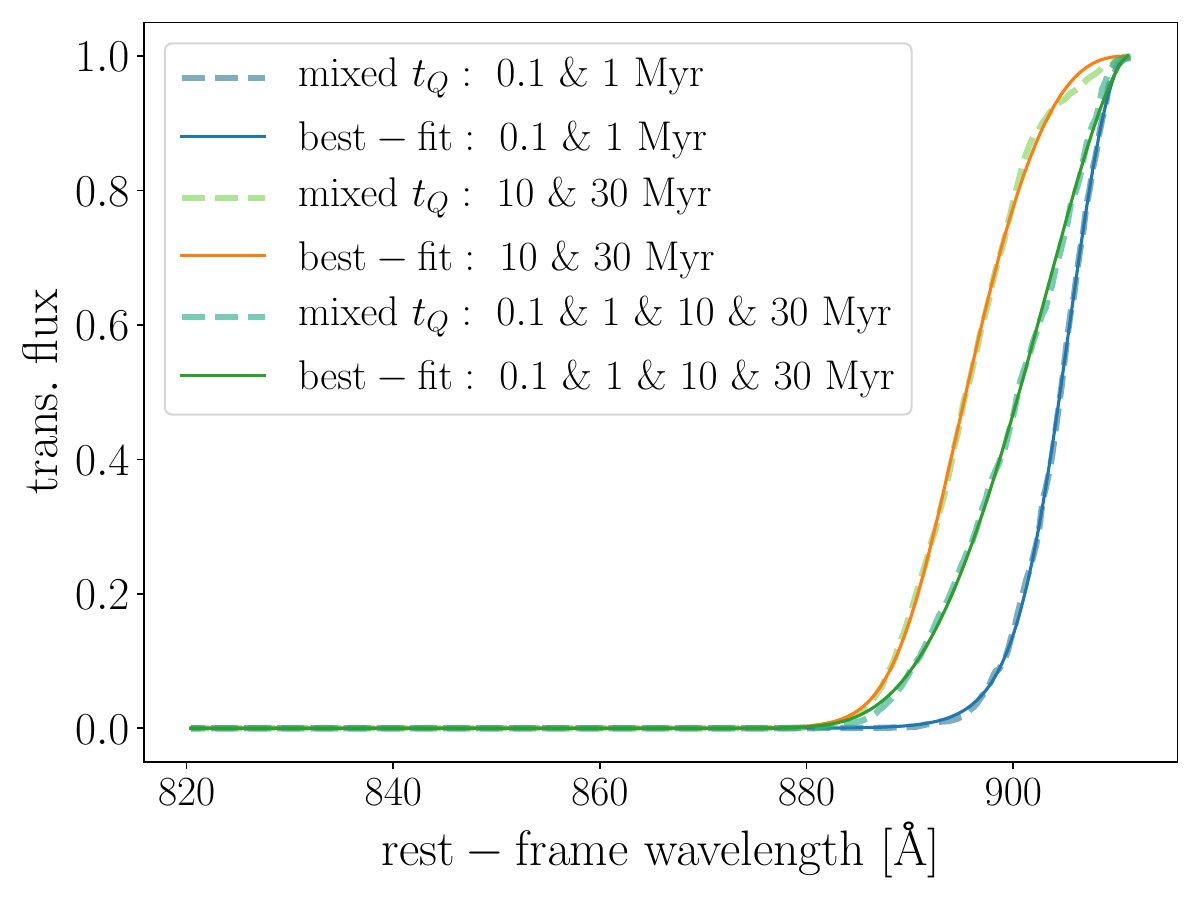}
    \caption{Mean profiles of mixed quasar lifetime $t_Q$ and the best-fit models for the $z=6.8$ snapshot. The blue dashed line are calculated with evenly mixed $t_Q=0.1~\rm~ Myr$ and $t_Q=1~\rm~ Myr$ profiles (500 each). The light-green dashed line is similar but evenly mixed with $t_Q=10~\rm~ Myr$ and $t_Q=30~\rm~ Myr$ profiles. The blue-green dashed line is an evenly mixture of all $t_Q$, 250 profiles each from $t_Q=0.1, 1, 10, 30 ~\rm~ Myr$. The solid lines are best-fit models with three fully-flexible parameters fit simultaneously.}
    \label{fig:fit_mixed_tq} 
\end{figure}



\subsection{{Fitting models to profiles with mixed $t_Q$}}

{True quasar lifetimes are highly uncertain \citep[e.g.,][]{satyavolu2023}. Given that the LyC absorption profile changes significantly with quasar lifetime for the $z=6.8$ snapshot, it would be meaningful to check the fitting result to the average profile with different $t_Q$. In this subsection, we specifically test fitting three parameters simultaneously, similar to the practice in \citet{becker2021,zhu2023}.}

{We perform the fitting to the following cases of samples of 1000 profiles\footnote{The values reported below are for the $\dot{N}=1.4\times 10^{57} ~\rm s^{-1}$ quasar profiles. Similar trends are also found for $\dot{N}=3.5\times 10^{56} ~\rm s^{-1}$ quasars.}: 1. mixture of small $t_Q$ profiles: we randomly draw 500 profiles of each $t_Q=0.1~\rm Myr$ and $t_Q=1~\rm Myr$,  and calculate the average profile, as shown by the blue dashed line in Figure \ref{fig:fit_mixed_tq}; 2. mixture of large $t_Q$ profiles: we randomly draw 500 profiles of each $t_Q=10~\rm Myr$ and $t_Q=30~\rm Myr$, and calculate the average profile, as shown by the light-green dashed line in Figure \ref{fig:fit_mixed_tq}; 3. mixture of all four $t_Q$: we randomly draw 250 profiles of each $t_Q=0.1, 1, 10, 30 ~\rm Myr$ and plot the average profile as the blue-green dashed line. Then we fit $\mfp$,$\Req$, $\xi$ simultaneously. The best-fit parameters and MSE for each case are listed in Table \ref{tab:mix_tQ_fit}. For Case 1, the best-fit $(\mfp \rm{[pMpc]}, \Req \rm{[pMpc]}, \xi)$= (1.2, 4.8, 0.50) and the MSE is $6.9\times 10^{-6}$, respectively. Compared with the true MFP ($0.4 ~\rm pMpc$) and $\Req$ ($36 ~\rm~ pMpc$), the best-fit results overestimates MFP and underestimates $\Req$. Similar trend is found for Case 2, where the best-fit and MSE are $(\mfp \rm{[pMpc]}, \Req \rm{[pMpc]}, \xi)$=
(1.0, 15, 0.83) and $3.0\times 10^{-5}$, respectively. As for Case 3, if we allow $\Req$ to be any positive number, then the best-fit routine returns unrealistically large $\Req$ -- the best-fit and MSE are $(\rm{[pMpc]}, \Req \rm{[pMpc]}, \xi)$ = (0.066, 13302, 0.26) and $2.6\times 10^{-5}$, respectively. We test setting a upper boundary for $\Req$, and find that the best-fit would always push $\Req$ to the upper limit. For example, setting the range of $\Req$ to be between 0 and $100$ pMpc leads to the best-fit and MSE to be (0.84, 100,0.26) and $2.7\times 10^{-5}$, respectively. Visually, both parameter combinations fit the profile well, showing a strong degeneracy between $\mfp$ and $\Req$. This indicates that in the regime of very small true MFP and a mean LyC profile that changes significantly with $t_Q$, fitting with fully flexible parameters could result in incorrect $\mfp$ in absence of a good prior for $\Req$.  }

\subsection{Comparison with previous studies}

Recently, \citet{satyavolu2024} and \citet{roth2024} have used different simulations to investigate the potential biases in the MFP measurements. \citet{satyavolu2024} uses a larger box of $160 \CHIMP$ by each side, where the most massive halo at $z=6$ is $\approx 4.6\times10^{12} \Msun$. The true MFP in their $z=6$ snapshot is $\gtrsim 2$ pMpc for both random sightlines and sightlines centered on massive halos. They have studied the LyC profile from halos $\sim 10^{12} \Msun$ at different $t_Q$, as well as sightlines centered on halos with different masses ranging from $10^{10} \sim 10^{12} \Msun$ with $t_Q$ fixed at 1 Myr. The snapshot with the closest MFP in our simulation is the $z=6.1$ one (middle panels in Fig. \ref{fig:profile_diff_tQ_all}).
However, due to the smaller box sizes in CROC, the masses of our 20 most massive halos ranges from $2-9 \times 10^{11} \Msun$.
The evolution of LyC with $t_Q$ is different from what they find for their $M_h\sim 10^{12} \Msun$ sightlines in several ways.

First, the difference of the mean LyC profile between our $t_Q=0.1$ Myr case and $t_Q = 1$ Myr cases is larger than that in \citet{satyavolu2024}, while the differences between our $t_Q=1$ Myr and $t_Q=10$ Myr cases are smaller.
Second, our profiles are smooth in the very inner part, while those in \citet{satyavolu2024} have a sharper decrease. These discrepancies could be a result from the differences in box sizes and spatial resolutions, as well as a result of different implementations of physics models. The peak resolution in CROC boxes is 100~pc, a higher resolution than that in \citet{satyavolu2024}. This could result in differences in the properties of (sub)-LLSs between the two simulations, which could significantly impact the LyC profile evolution on timescales $\lesssim 0.1$ Myr. On the other hand, due to the larger volume of their box, there are {likely to be more}  {coherent} neutral patches in the boxes used in \citet{satyavolu2024}. {In other words, the neutral patches might be larger and coherent in their large box while smaller and shattered in our small box. Large coherent neutral patches could cause slower propagation of quasar I-front systematically. It could potentially explain why the average LyC profile in \citet{satyavolu2024} (see their Fig. 3) exist a sharp decrease in the inner part. We see a similar feature in snapshot with neutral patches (left panels in  \ref{fig:profile_diff_tQ_all}), but it is less sharp because the neutral patches distributed in a smaller box is typically smaller and more shattered. As a result, the quasar I-front will slow down for a shorter time at very  different location for different sightlines, thus a sharp drop feature do not show up  in an averaged profile. }

On the other hand, the \citet{roth2024} analysis uses sightlines drawn from  a small box of $3 \CHIMP$ on each side with a resolution of $N=2\times 512^3$ particles. This small box simulation is fully ionized with uniform background. To mimic a not-fully ionized universe, they modify the neutral fraction to unity in selected regions. This reassignment of neutral fraction is done by utilizing sightlines drawn from a separate large-volume low-resolution simulation with box size $(200 \CHIMP)^3$ and $200^3$ RT grid cells. They identify the neutral IGM sections in these set of sightlines and switch the neutral fraction to one at the same distance in the original set of sightlines.
In other words, the density and neutral fraction in the sightlines are decoupled. Using this simulation, they test the case where the true MFP $\approx 1$ pMpc. They find that the inferred MFP depends on quasar lifetime at $<50\%$ level. This level of bias/uncertainty is consistent with what we have found for snapshots where the true MFP is $\gtrsim 2$~pMpc.
However, \citet{roth2024} does not have an explicit test of the scenario where the true MFP $\lesssim 0.5$ pMpc. 
For this scenario, it is likely that the level of uncertainty in the inferred MFP from the \citet{roth2024} might be also be consistent with what we have found given the consistency with the larger true MFP scenarios.

\subsection{Implication for quasar lifetime constraints}\label{sec:quasarlifetime}

\begin{figure}\label{fig:Lya}
    \centering
    \includegraphics[width=0.45\textwidth]{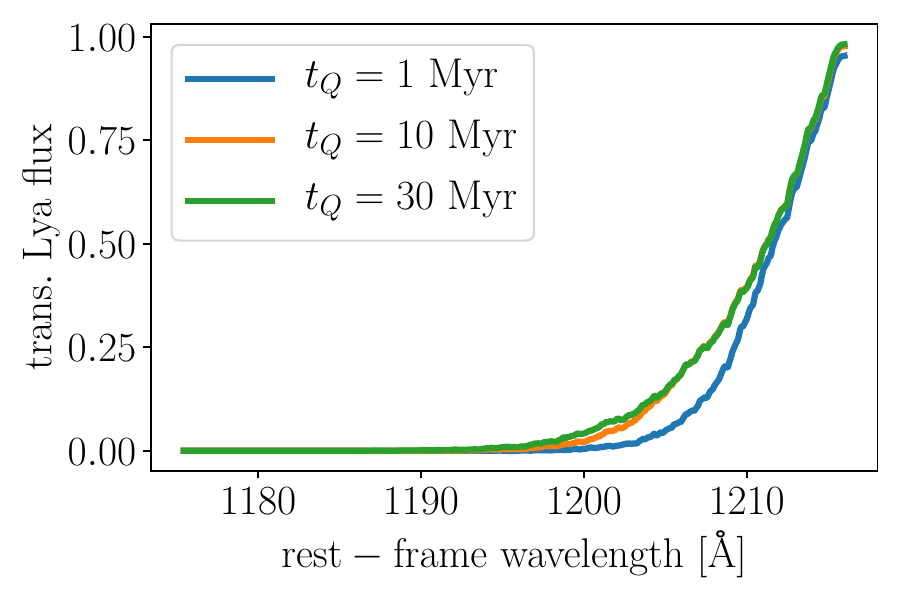}
    \caption{The mean Ly$\alpha$ profiles at different $t_Q$ for snapshot $z=6.8$.}
    \label{fig:Lya}
\end{figure}

Recovering the true MFP can be challenging in the regime when there are neutral patches in the universe and the true MFP is short ($\lesssim 0.5$ pMpc). On the other hand, the LyC profile has significant dependence on the quasar lifetime in this regime (see strong dependence on quasar lifetime in the left most panels of Figure~\ref{fig:profile_diff_tQ_all} when significant neutral patches exist). This regime could therefore provide a powerful testing ground to constrain quasar lifetimes. The Ly$\alpha$ proximity zone has been used to constrain quasar lifetime \citep[e.g.,][]{eilers2017, morey2021}. However, Lya proximity zone is only sensitive to timescales shorter than $t_Q<10^5$ yr. Indeed, in Figure~\ref{fig:Lya}, we show the mean Ly$\alpha$ profiles of different quasar lifetimes. The profile shapes of the cases where $t_Q>1$ Myr are almost identical, while the difference between $t_Q=1$ and $t_Q=10$ yr profiles is only apparent in the outer part of the spectra. On the contrary, the difference between these timescales is much more obvious in the LyC spectra, due to the lower cross section of LyC. Finally, the LyC profile could be a better tracer than the Ly$\alpha$ profile of the cumulative quasar lifetime because the LyC profile is sensitive to the quasar lifetime on much longer timescales. The Ly$\alpha$ profile is a better tracer for the episodic quasar lifetime within a few thousand years.

However, we point out the practical difficulties in using the LyC to constrain quasar lifetimes.
First of all, LyC is much harder to measure than Ly$\alpha$ due to the foreground Lyman series absorption.
The Lyman series has wavelengths $\frac{n^2}{n^2-1}$ of the Lyman limit, where $n$ is the level of prime energy state.
This means that at the Lyman edge of a $z=6$ quasar, the Ly$\alpha$ series at $z=4.25$, the Ly$\beta$ at $z=5.2$ and the higher-order series at higher redshifts all create foreground absorption.
The neutral fraction above $z=4$ is significant enough to have an effective Ly$\alpha$ optical depth $\approx 1$ \citep{bolton2017}.
This adds difficulties in extracting the intrinsic LyC spectra. In general, quasars at $z_{\rm qso}$ suffer from foreground absorption of Lyman series $n$ at $z_{\rm fg}=\frac{n^2-1}{n^2} z_{\rm qso}-\frac{1}{n^2}$. Because the Lyman series opacity is higher at higher $z_{\rm fg}$, higher redshift quasars suffer from a correspondingly higher level of foreground absorption. The degree to which we can separate the effects of foreground Lyman series absorption for higher redshift quasars is underexplored. It is a future direction of research to combine the information from the intrinsic LyC and Lyman series proximity zones to constrain quasar lifetimes on different timescales.

\section{Conclusion}
\label{sec:concl}
We study the LyC absorption spectra by post-processing CROC simulations. We choose three snapshots around the ``ankle point'' of the box reionization history, where all neutral patches have just disappeared in the box. The true MFP of the box are $\approx 0.4, 2, 6$~pMpc at $z\approx 6.8, 6.1, 5.4$, respectively. Our key findings are as follows:

\begin{itemize}
    \item In the very small MFP regime ($\approx 0.4$~pMpc), we find that the mean LyC absorption profile is very sensitive to quasar lifetime on all timescales. The slow evolution of LyC profile under quasar radiation is due to a combination of the extended neutral IGM patches and the dense neutral clumps that block the ionizing radiation of the quasar. In this regime, the measured MFP using \B21 procedure can be a factor of few different from the true MFP depending on the quasar lifetime.
    
    \item In the mid MFP regime ($\approx 2$~pMpc), the mean LyC profile does not evolve significantly after $t_Q \sim 1$ Myr. However, the numerous dense neutral clumps shield quasar radiation and cause significant differences between the profile at $t_Q= 1$~Myr and $t_Q= 0.1$~Myr. For $t_Q\gtrsim 1$~Myr, the best fit $\mfp$ is biased high by $\approx 30\%$ if the correct $\Req$ and the fiducial $\xi=0.67$ is used.
    
    \item In the long MFP regime ($\approx 6$~pMpc), the mean LyC profile stabilizes after the quasar has been shining for $t_Q \sim 1$ Myr. The difference between $t_Q\gtrsim 1$~Myr and $t_Q\sim 0.1$~Myr is also minimal, due to significantly fewer neutral clumps in the IGM. The best fit $\mfp$ is biased high by $\approx 20\%$ if the correct $\Req$ and the fiducial $\xi=0.67$ is used.
    
    \item The sample variance for $10$ quasar sightlines is relatively small, with a relative standard deviation $15\%-30\%$ in different scenarios.
    
    \item We simultaneously fit the three free parameters and find a similar $\mfp$, albeit slightly larger, to fits with $\Req$ and $\xi$ fixed in cases where the true MFP $\gtrsim 2$ pMpc.
\end{itemize}

For snapshots where the true MFP is $\gtrsim 2$ pMpc, our conclusions are consistent with previous findings \citep{satyavolu2024,roth2024}. However, when the MFP reaches very small regime ($\lesssim 0.5$ pMpc), we caution the use of the procedure in \citet{becker2021}, as it may result in unphysical results. The regime of MFP$\lesssim 0.5$ pMpc, however, can be highly valuable for constraining the cumulative lifetime of the first quasars, which we will explore in future studies.

\section*{Acknowledgments}
The authors are grateful to Nick Gnedin and Hanjue Zhu for invaluable discussions.
The authors thank Yongda Zhu and Christopher Cain for productive discussions.
HC thanks the support by the Natural Sciences and Engineering
Research Council of Canada (NSERC), funding reference \#DIS-2022-568580, \#RGPIN-2025-04798, \#DGECR-2025-00136 and by the University of Alberta, Augustana Campus Department of Science. HC also thanks the ``Cosmic Dawn at High Latitudes Conference'' sponsored by NORDITA for helpful discussions. CA acknowledges
support from DOE grant DE-SC009193, and the Leinweber
Center for Theoretical Physics.

\bibliographystyle{apsrev4-1}

\bibliography{oja_template}

\end{document}